\documentclass[fleqn,usenatbib]{mnras}

\usepackage{mathptmx}

\usepackage[T1]{fontenc}

\DeclareRobustCommand{\VAN}[3]{#2}
\let\VANthebibliography\thebibliography
\def\thebibliography{\DeclareRobustCommand{\VAN}[3]{##3}\VANthebibliography}

\usepackage{graphicx}	
\usepackage{amsmath}	
\usepackage{amssymb, mathtools}	
\usepackage{natbib, url}
\usepackage[dvipsnames]{xcolor}

\newcommand{\Msol}{\rm{M_{\odot}}}

\newcommand{\kms}{\rm{km\,s^{-1}}}
\newcommand{\Msolyr}{\Msol\,yr^{-1}}

\newcommand{\cm}{{\rm cm}}

\newcommand{\Tstar}{T_\star}

\newcommand{\Mdot}{\dot{M}}

\newcommand{\Rstar}{R_\star}

\newcommand{\vexp}{v_{\rm exp}}
\newcommand{\eps}{\varepsilon}

\title[Sensitivity study of chemistry in AGB outflows]{Sensitivity study of chemistry in AGB outflows using chemical kinetics}

\author[S. Maes et al.]{S.\ Maes$^{1}$\thanks{E-mail: {silke.maes@kuleuven.be}}, 
	M.\ Van de Sande$^{2}$, T.\ Danilovich$^{3,1}$, F.\ De Ceuster$^{1}$ \& L.\ Decin$^{1,4}$
	\\
	$^{1}$ Institute for Astronomy, KU Leuven, Celestijnenlaan 200D,
	3001 Leuven, Belgium\\
	$^{2}$ School
	of Physics and Astronomy, University of Leeds, Leeds LS2 9JT, UK
	\\
	$^{3}$ 
	School of Physics \& Astronomy, Monash University, Wellington Road, Clayton 3800, Victoria, Australia
	\\
	$^{4}$ School of Chemistry, University of Leeds, Leeds LS2 9JT, UK
}

\date{Accepted 2023 April 8. Received 2023 March 30; in original form 2022 October 24}

\pubyear{2023}

\begin{document}
	\label{firstpage}
	\pagerange{\pageref{firstpage}--\pageref{lastpage}}
	\maketitle
	
	\begin{abstract}
	Asymptotic Giant Branch (AGB) stars shed a significant amount of their mass in the form of a stellar wind, creating a vast circumstellar envelope (CSE). Owing to the ideal combination of relatively high densities and cool temperatures, CSEs serve as rich astrochemical laboratories. While the chemical structure of AGB outflows has been modelled and analysed in detail for specific physical setups, there is a lack of understanding regarding the impact of changes in the physical environment on chemical abundances. A systematic sensitivity study is necessary to comprehend the nuances in the physical parameter space, given the complexity of the chemistry. This is crucial for estimating uncertainties associated with simulations and observations. In this work, we present the first sensitivity study of the impact of varying outflow densities and temperature profiles on the chemistry. With the use of a chemical kinetics model, we report on the uncertainty in abundances, given a specific uncertainty on the physical parameters. Additionally, we analyse the molecular envelope extent of parent species and compare our findings to observational studies. Mapping the impact of differences in physical parameters throughout the CSE on the chemistry is a strong aid to observational studies.
	\end{abstract}
	
	\begin{keywords}
		astrochemistry -- molecular processes -- stars: AGB and post-AGB -- circumstellar matter -- ISM: molecules
	\end{keywords}
	
	
	
	\section{Introduction}
	AGB stars are evolved, low- to intermediate-mass stars ($\sim 0.8 - 8\, \Msol$) that are characterised by significant mass loss due to a dust-driven wind \citep{Bowen1988}. As a result, this outflow creates a vast circumstellar envelope (CSE). The mass-loss rates of AGB stars range from $\sim10^{-8}$ to $10^{-4}\,\Msolyr$ and expansion velocities are typically found to vary between $3$ and $30$\,$\kms$ \citep{Knapp1998,Habing2004agbs.book,Ramstedt2009,HofnerOlofsson}. The outflows of AGB stars are rich in chemistry, where over 100 molecules and about 15 dust species have been detected so far (e.g., \citealp{Habing1996,Verhoelst2009, Gail2013,Decin2021}). Their chemical richness is thanks to the large gradients in density and temperature throughout the outflow. The type of chemistry in the CSE is set by the elemental carbon-to-oxygen (C/O) ratio of the AGB star, with ${\rm C/O}<1$ resulting in oxygen-rich outflows and ${\rm C/O}>1$ resulting in carbon-rich outflows. AGB outflows with ${\rm C/O}\approx1$ are referred to as S-type. The CSE consists out of three regions, each characterised by specific physical and chemical conditions. In the inner wind ($\sim 1-5\ \Rstar$), the chemistry is taken out of equilibrium due to shocks resulting from pulsations emerging at the stellar surface. This leads to the presence of C-rich molecules in O-rich outflows, such as HCN, and vice versa, such as H$_2$O (e.g., \citealp{Bujarrabal1994,DecinAgundez2010}). In the intermediate region ($\sim 5-100\ \Rstar$), dust grains are able to grow and consequently launch the outflow. The outer wind ($\sim 100-20\,000\ \Rstar$) is dominated by photochemistry driven by interstellar UV photons penetrating the CSE (e.g., \citealp{HofnerOlofsson}). 
	\\ \indent The abundance, or even the mere presence, of different chemical species provides a powerful tool to probe the physical properties of CSEs and to study its kinematics and morphology. Moreover, through heating and cooling processes, the specific composition of species feeds back into the CSE structure (e.g.,  \citealp{Sahai1990,Decin2006}). Hence, accurate knowledge of the chemistry within the outflow and of how it depends on the physical conditions is crucial to our understanding of AGB outflows.
	\\ \indent Chemical models of outflows of individual AGB sources have been calculated, assuming specific physical conditions retrieved from observations  (e.g.\, \citealp{MH1994,WillacyMillar1997,Cherchneff2006,Li2016,Agundez2020}). Additionally, theoretical studies have been carried out about the effects on the chemistry due to, e.g., deviations from spherical symmetry of CSEs, dust-gas interactions, and companion photons (e.g., \citealp{VdS2018, VandeSande2019,VdSMillar2022}). These studies provide us with new insights on the complexity of the chemistry when compared to observations.
	\\ \indent However, up until now, the effect of changes in the physical environment of the CSE (such as its density and temperature) on the chemistry, remains largely unknown. {This is a missing piece of work, since, given the complexity of these models, the resulting abundances can depend on these physical parameters in a non-trivial way. The impact of changes in the physical parameter space are key to estimate uncertainties on theoretical as well as on observational results. Therefore, a sensitivity analysis is a crucial step in establishing substantiated confidence in the predictions of these models. Moreover, it }will benefit observational studies in constraining the system parameters and dynamics of the outflow (e.g., \citealp{Danilovich2016}).
	\\ \indent We present the first sensitivity study of CSE chemistry to the underlying physical environment. More specifically, we examine the impact of altering the temperature profile and the outflow density of the CSE on the chemical abundances and molecular envelope sizes. {We derive a theoretical uncertainty on these abundances, given a specific uncertainty on the physical parameters. For this,} we use a chemical kinetics approach, because of the non-equilibrium setting of the outflow.
	\\ \indent This paper is organised as follows. The modelling setup and the parameter space of the studied grid is introduced in Sect.\ \ref{sect:method}. Sect.\ \ref{sect:results} presents the abundance profiles of different species from selected models. In Sect.\ \ref{sect:discussion}, we analyse and discuss the chemistry behind the variation in the abundance profiles. {We elaborate on envelope sizes and reflect on the approximation made for the CO self-shielding in our chemical model. In Sect.\ \ref{sect:sens}, the sensitivity of the chemistry to the physical parameters is discussed. In Sect.\ \ref{sect:observations}, we compare our molecular envelope sizes to} observational studies. In Sect.\ \ref{sect:conclusion}, we summarise and conclude.

	\section{Methodology}\label{sect:method}
	\subsection{Chemical model of the circumstellar envelope}\label{sect:chem_model}
	The chemical kinetics model used is based on the publicly available one-dimensional CSE model of the UMIST Database for Astrochemistry (UDfA, \citealp{McElroy2013}\footnote{\label{footnote:UDfA} \url{http://udfa.ajmarkwick.net/index.php?mode=downloads}}), which computes the abundances of chemical species as a function of distance from the star.
	\subsubsection{Physics}
	The model assumes a smooth spherically symmetric outflow with constant expansion velocity, $\vexp$, and mass-loss rate, $\Mdot$. Therefore, the gas density falls as $1/r^2$, given by
	\begin{equation}\label{eq:density}
		\rho(r) = \frac{\Mdot}{4\pi r^2 \vexp\, \mu\, m_{\rm H}},
	\end{equation}
	with $\mu$ the mean molecular mass per H$_2$ molecule and $m_{\rm H}$ the atomic mass unit.
	The kinetic temperature profile throughout the outflow is governed by a power-law with exponent $\varepsilon$, implemented by \cite{VdS2018}:
	\begin{equation}\label{eq:temp_profile}
	T = \Tstar \left(\frac{r}{\Rstar}\right)^{-\varepsilon},
	\end{equation} 
	where $\Tstar$ is the surface temperature of the AGB star and $\Rstar$ the stellar radius. Such a profile has proven to be a good representation for the kinetic temperature in CSEs \citep{Millar2004}. A lower limit of 10\,K is imposed on the temperature profile to avoid unrealistic temperatures in the outer parts of the CSE \citep{CM2009}. H$_2$ is assumed to be completely self-shielded, CO self-shielding is implemented using the single-band approximation from \cite{MJ1983}. Further details about the model can be found in \cite{Millar2000}, \cite{CM2009}, and \cite{McElroy2013}.
	
	\subsubsection{Chemistry}\label{sect:kinetics}
	Chemical abundances are computed using chemical kinetics, i.e.\ by solving a set of coupled ordinary differential equations representing the evolution of the number density of each species. The change in number density $n_i$ of species $i$ is given by 
	\begin{equation}\label{eq:ODE}
		\frac{{\rm d}n_i}{{\rm d}t} = \sum_{j,l}k_{jl}n_jn_l + \sum_{m}k_mn_m-n_i\left(\sum_rk_{ir}n_r+\sum_sk_s\right) \quad [{\rm cm^{-3}\, s^{-1}}],
	\end{equation}
	since the chemical network only comprises one- and two-body reactions. The first two terms give the rate of change due to formation reactions and the two last terms due to destruction reactions of species $i$, $k$ is the rate coefficient for the specific reaction. The one-body reactions (rate coefficients $k_{m}$ and $k_{s}$ with units of ${\rm s^{-1}}$ in Eq.\ \ref{eq:ODE}) consist of photodissociation reactions by interstellar or cosmic ray photons, and photoionisation. For two-body reactions, the rate coefficient is parametrised using the modified Arrhenius equation ($k_{jl}$ and $k_{ir}$ in Eq.\ \ref{eq:ODE}):
	\begin{equation}\label{eq:rate_coeff}
	k= \alpha \left( \frac{T}{300\,{\rm K}}\right)^{\beta}\exp\left(\frac{-\gamma}{T}\right) \quad [{\rm cm^3\, s^{-1}}],
	\end{equation}
	where the constants $\alpha$, $\beta$, and $\gamma$ belong to a particular reaction: $\alpha$ is given in ${\rm cm^3\, s^{-1}}$, $\beta$ indicates the temperature dependence, and $\gamma$ (given in K) the energy barrier $\gamma k_B$, with $k_B$ the Boltzmann constant. If $\beta$ and/or $\gamma$ differ from zero, the reaction rate will be dependent on temperature. The two-body reactions include reactions between neutral species, but also with ions and electrons. Reactions with ions (e.g., ion-neutral or mutual neutralisation) often have $\beta = -0.5$ \citep{Smith2011}, resulting in an inverse dependency on the temperature. Hence, the reaction rates increase for decreasing temperature (Eq.\ \ref{eq:rate_coeff}). For dissociative recombination and radiative association reactions, $\beta$ often even takes larger negative values of order unity. On the other hand, reactions between neutral species are faster at higher temperatures, thus $\beta > 0$. The latter reactions also exhibit an energy barrier $\gamma k_B$ when no radicals are involved, with $\gamma$ being several hundreds to thousands of Kelvin, due to the electronic rearrangement of the newly formed molecule. 
	\\ \indent The chemical network used is based on \textsc{Rate12}, the most recent release of the UDfA$^{\ref{footnote:UDfA}}$ \citep{McElroy2013}. It consists of gas-phase chemistry only, involving 467 different species connected by 6173 reactions. 
	All abundances are initially set to zero, except for a set of parent species. These are assumed to have formed in the inner wind and injected into the CSE at a radius of $10^{14}\,\cm$, the starting radius of our models. We consider both an O-rich and C-rich CSE. The sets of parent species are based on observational studies and are taken from \cite{Agundez2020}. They are listed in Table \ref{tab:parents}. 
	\begin{table}
		\begin{center}
			\caption{Parent species of the C-rich and O-rich outflows, and their initial abundances relative to H$_2$.}
			\begin{tabular}{   l   r   l  r       }
				\hline \hline \\[-2ex]
				Carbon-rich &  & Oxygen-rich & \\ 
				{Species} & {Abundance } & 	{Species} & {Abundance }  \\ \hline
				He          & $0.17$ & He       & $0.17$  \\ 
				CO          & $8.00\times10^{-4}$ & CO       & $3.00\times10^{-4}$  \\ 
				C$_2$H$_2$  & $4.38\times10^{-5}$ & H$_2$O   & $2.15\times10^{-4}$  \\ 
				HCN         & $4.09\times10^{-5}$ & N$_2$    & $4.00\times10^{-5}$  \\ 
				N$_2$       & $4.00\times10^{-5}$ & SiO      & $2.71\times10^{-5}$  \\ 
				SiC$_2$     & $1.87\times10^{-5}$ & H$_2$S   & $1.75\times10^{-5}$  \\ 
				CS          & $1.06\times10^{-5}$ & SO$_2$   & $3.72\times10^{-6}$  \\ 
				SiS         & $5.98\times10^{-6}$ & SO       & $3.06\times10^{-6}$  \\ 
				SiO         & $5.02\times10^{-6}$ & SiS      & $9.53\times10^{-7}$  \\ 
				CH$_4$      & $3.50\times10^{-6}$ & NH$_3$   & $6.25\times10^{-7}$  \\ 
				H$_2$O      & $2.55\times10^{-6}$ & CO$_2$   & $3.00\times10^{-7}$  \\ 
				HCl         & $3.25\times10^{-7}$ & HCN      & $2.59\times10^{-7}$  \\ 
				C$_2$H$_4$  & $6.85\times10^{-8}$ & PO       & $7.75\times10^{-8}$  \\ 
				NH$_3$      & $6.00\times10^{-8}$ & CS       & $5.57\times10^{-8}$  \\ 
				HCP         & $2.50\times10^{-8}$ & PN       & $1.50\times10^{-8}$  \\ 
				HF          & $1.70\times10^{-8}$ & F        & $1.00\times10^{-8}$  \\ 
				H$_2$S      & $4.00\times10^{-9}$ & Cl       & $1.00\times10^{-8}$  \\ 
				
				 \hline						
			\end{tabular}
			\label{tab:parents}
		\end{center}
		{\footnotesize \textbf{Notes.} Abundances taken from \cite{Agundez2020}. When a range was given there, the linear average is used.}
	\end{table}
	\\ \indent The chemistry in the outer wind is triggered by photodissociation of the parent species, resulting in a cascade of chemical reactions \citep{Saberi2019IAUS,Millar2020}. This is caused by the interstellar UV radiation field, {for which we use interstellar radiation field from} \cite{Draine1978}. The UV field is extinguished by dust, using the approach of \cite{JM1981}. 
	\begin{figure*}
		\centering
		\includegraphics[width=0.4\textwidth]{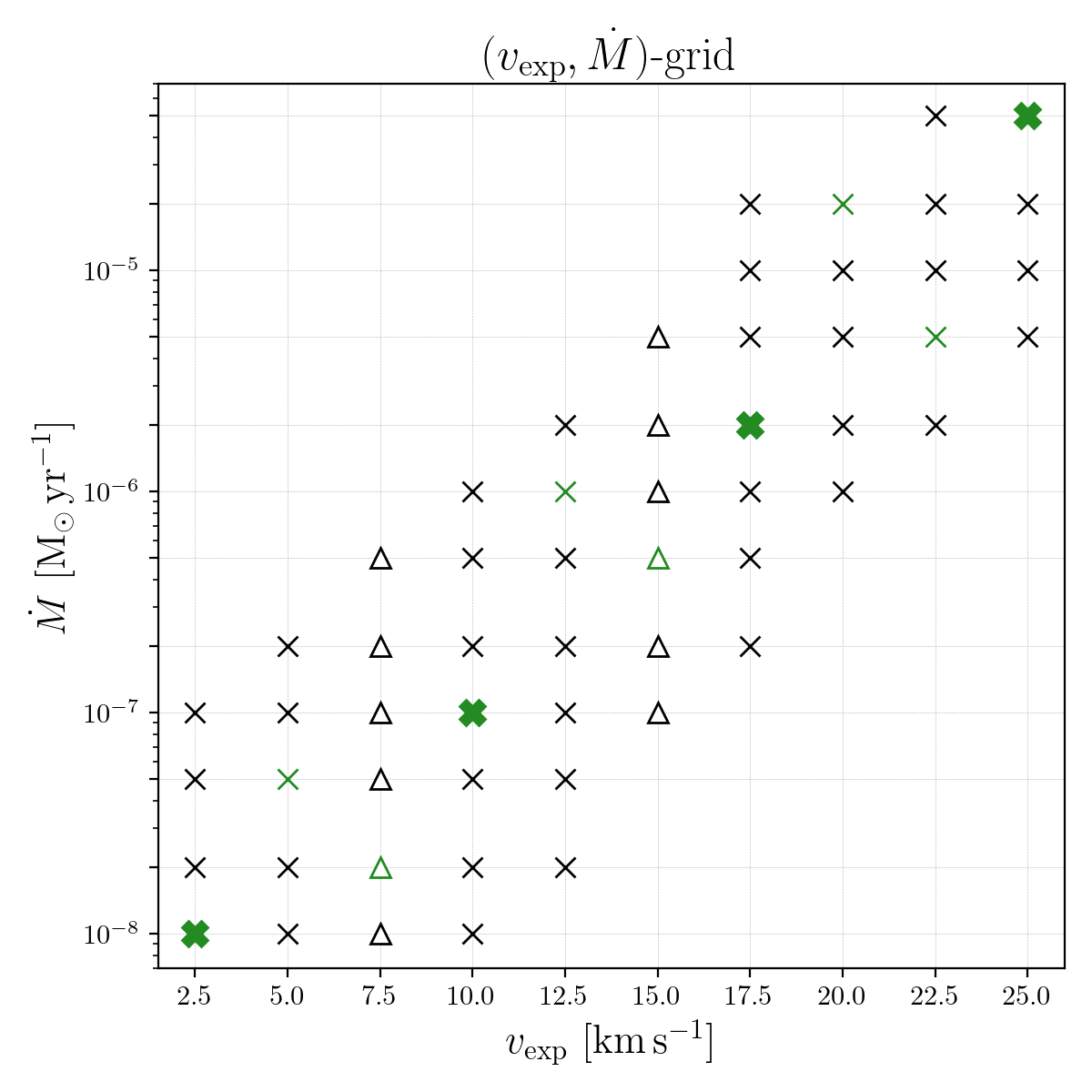}
		\hspace{1.5cm}
		\includegraphics[width=0.425\textwidth]{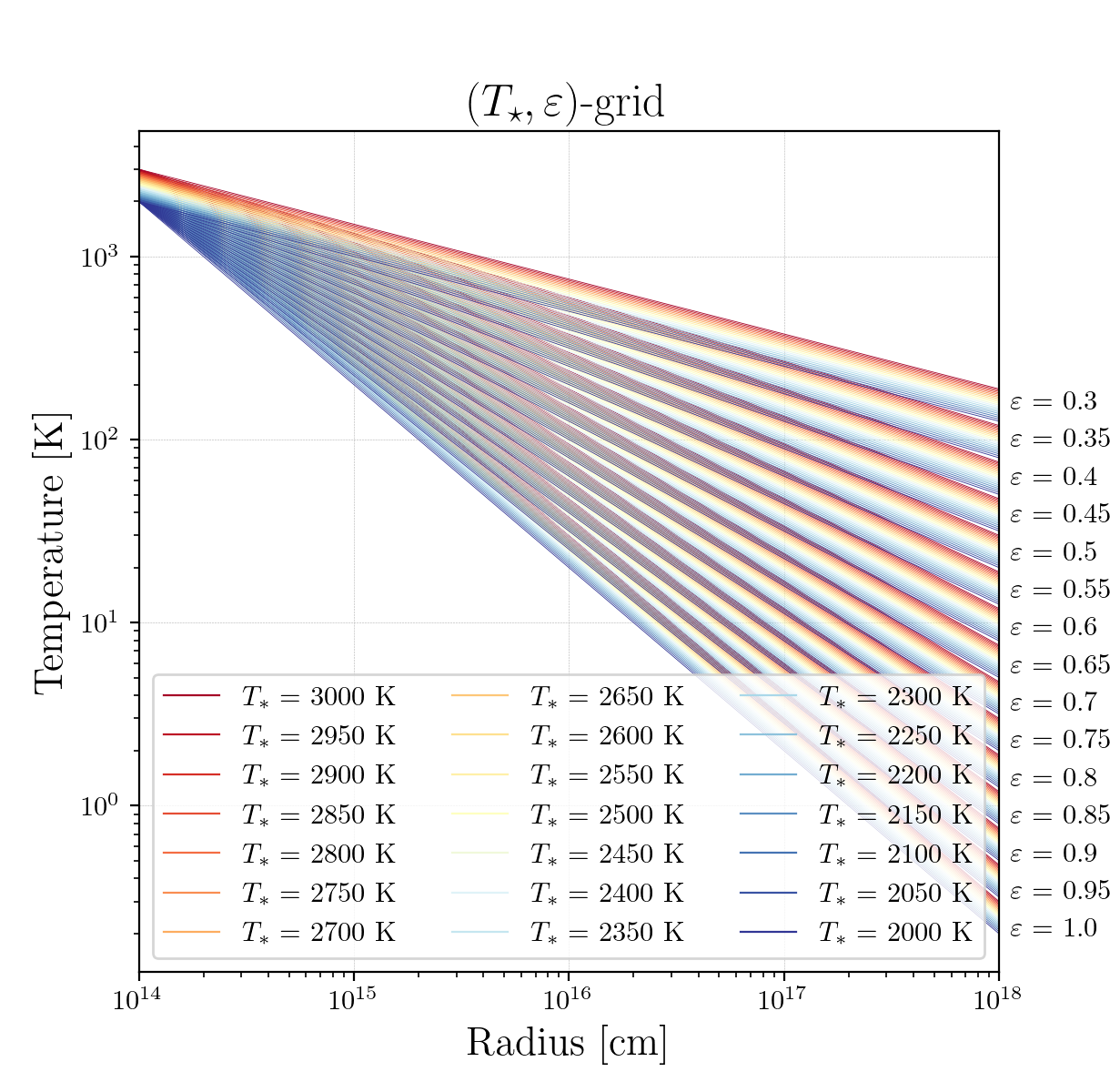}
		\caption{{Visualisation of the parameter space of our grid. \textit{Left}: The symbols show the different outflow densities via the combinations of expansion velocity, $\vexp$, and mass-loss rate, $\Mdot$ (Eq.\ \ref{eq:density}). The bold crosses highlight the modelled densities displayed in Sect.\ \ref{sect:results} and \ref{sect:discussion}, the symbols highlighted in green are used for the sensitivity study in Sect.\ \ref{sect:sens}, and the triangles indicate the models overlapping with \protect\cite{Saberi2019} and \protect\cite{Groenewegen2017} (Sect.\ \ref{sect:COselfshielding}). \textit{Right}: Visualisation of the different temperature profiles (Eq.\ \ref{eq:temp_profile}), where the stellar temperature, $\Tstar$, is indicated by the line colour. The different values of $\eps$ result in different groups of temperature profiles, indicated at the right-hand side of the panel.}}
		\label{fig:grid}
	\end{figure*}

	\subsection{Physical parameter space}
	To determine the sensitivity of chemical abundances to the physical environment of the CSE, we vary the outflow density and the temperature profile. An overview of the grid parameters is given in Table \ref{tab:grid} and visualised in Fig.\ \ref{fig:grid}.
	\\ \indent The density of the outflow is determined by the mass-loss rate, $\Mdot$, and the expansion velocity, $\vexp$ (Eq.\ \ref{eq:density}). Hence, we constructed a grid by changing the combination of $\Mdot$ and $\vexp$.	The mass-loss rate is varied between $1\times10^{-8}$ and $5\times10^{-5}$ $\Msolyr$ and the expansion velocity ranges from 2.5 to 25 $\kms$ in steps of $2.5\,\kms$.  Observationally a linear correlation has been found between $\vexp$ and $\Mdot$, where generally higher expansion velocities go together with higher mass-loss rates (e.g., \citealp{Ramstedt2009}). This was taken into account when constructing the grid by excluding combinations of high mass-loss rate with small expansion velocity, and vice versa. The resulting 54 combinations of $(\vexp,\Mdot)$ are visualised in the left panel of Fig.\ \ref{fig:grid}.
	\\ \indent The temperature profile of the CSE is set by stellar temperature $\Tstar$ and the exponent $\eps$ (Eq.\ \ref{eq:temp_profile}). Different values are taken for $\Tstar$ and $\eps$, where high values of $\eps$ result in a steep temperature profile and low values give a more gradual profile, hence resulting in an overall warmer CSE. Generally, from observations $\eps$ is found to be around 0.6-0.7 \citep{Millar2004,DeBeck2010,Maercker2016}. Therefore, to fully explore the parameter space, we varied $\eps$ from 0.3 to 1.0 with intervals of 0.05. The stellar temperature ranges from 2000 to 3000\,K with intervals of 50\,K. This gives 315 different temperature profiles, visualised in the right panel of Fig.\ \ref{fig:grid}. In total, we calculated 17\,010 O-rich and 17\,010 C-rich models, giving of 34\,020 models overall.
	\begin{table}
		\begin{center}
			\caption{Physical parameters of the grid, together with the ranges and stepsizes, if variable. The density is determined by $\Mdot$ and $\vexp$ according to Eq.\ (\ref{eq:density}) for the combinations given in Fig.\ \ref{fig:grid}, the temperature profile is given by the combination of $\Tstar$ and $\eps$ through Eq.\ (\ref{eq:temp_profile}). The stellar radius, $\Rstar$, inner radius, $R_{\rm inner}$, and outer radius, $R_{\rm outer}$, are kept constant.}
			\begin{tabular}{   l l   c   c    }
				\hline \hline \\[-2ex]
				Parameter & & Range/Value & Stepsize   \\ \hline
				$\Mdot$ &[$\Msolyr$] &$1\times10^{-8}$ -- $5\times 10^{-5}$ & (*) \\
				$\vexp$ &[$\kms$] & 2.5 -- 25 & 2.5	 \\ 
				$\Tstar$ &[K] & 2000 -- 3000 & 50 \\ 
				$\varepsilon$& / & 0.3 -- 1.0 & 0.05  \\ \hline
				$\Rstar$ & [cm] & $2\times 10^{13}$ & / \\
				$R_{\rm inner}$ & [cm] & $10^{14}$ & /  \\ 
				$R_{\rm outer}$ & [cm] & $10^{18}$ & / 
				\\ \hline
			\end{tabular}
			\label{tab:grid}
		\end{center}
		{\footnotesize \textbf{Note.} (*) For $\Mdot$ we used the values $1\times10^{-p}$, $2\times10^{-p}$ and $5\times10^{-p}$, with $p\in[5,8]$, see also Fig.\ \ref{fig:grid}.  
		}
	\end{table}

	\section{Results}\label{sect:results}
	We investigated the effect of different outflow densities and different temperature profiles on the chemistry throughout the CSE. We consider its effects on the shape of the abundance profiles of specific parent and daughter species. To visualise the different temperature profiles, we colour-coded the abundance profiles using a ``reference temperature'': we use the temperature at $r = 10^{18}\,{\rm cm}$, following Eq.\ (\ref{eq:temp_profile}), so that there is no degeneracy in colour for each $(\Tstar,\eps)$-pair. Consequently, this reference temperature has no physical meaning and only indicates the steepness of the temperature profile within the modelled outflow. 
	\\ \indent In this section we show and consider how the abundance profiles change due to different outflow densities and temperature profiles for both O-rich and C-rich outflows. In Sect.\ \ref{sect:var_in_abs} we elaborate on the cause of the changes.

	\subsection{O-rich outflows}\label{sect:results_O}
	\begin{figure*}
		\centering
		\includegraphics[width=0.95\textwidth]{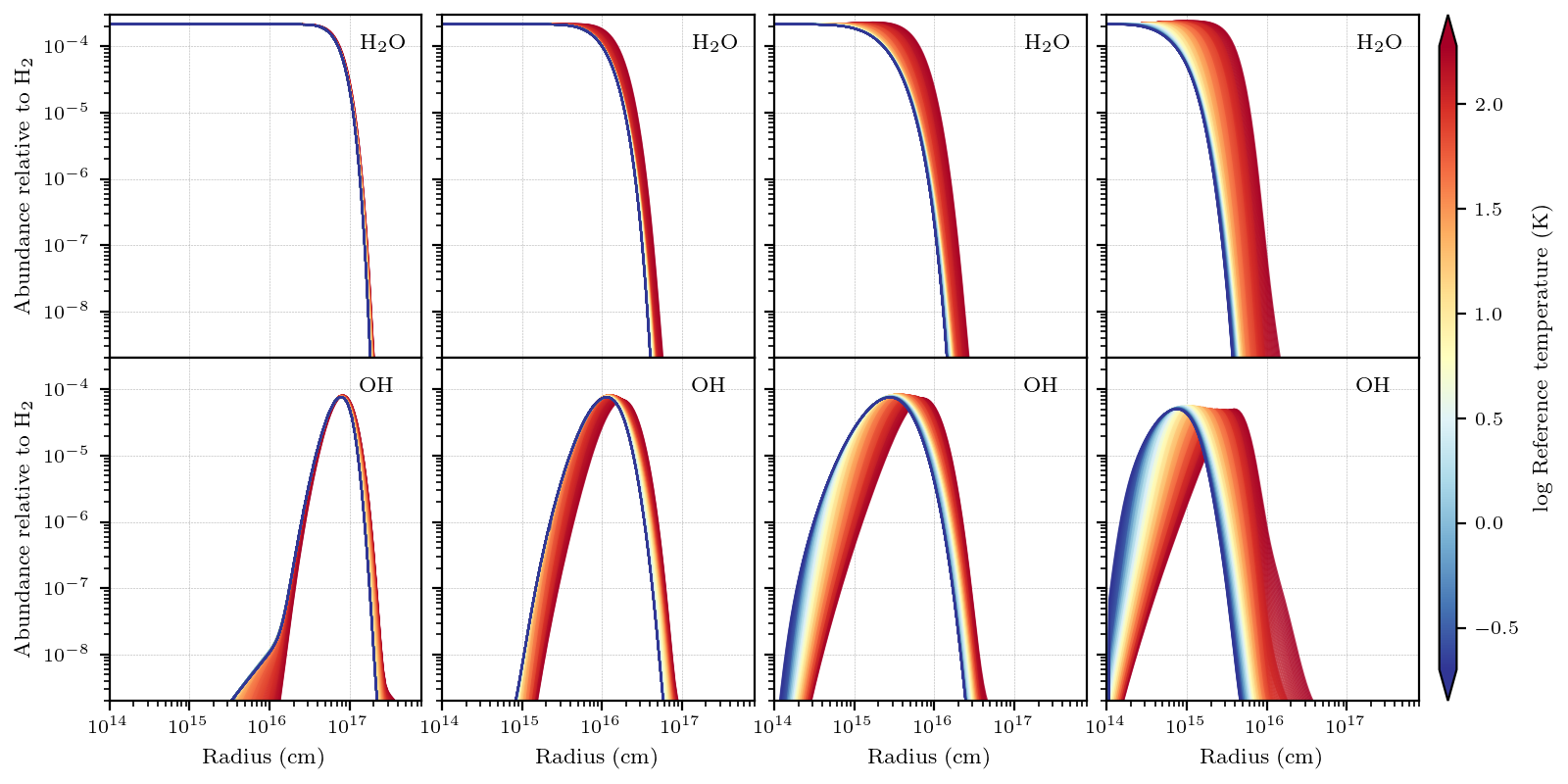}
		\caption{Fractional abundance profiles of H$_2$O and OH in O-rich outflows, going from high to low outflow density. From \textit{left} to \textit{right} $(\vexp \ [{\rm \kms}],\Mdot \ [{\rm \Msolyr}])$: (25.0,\ $5\times10^{-5}$), (17.5,\ $2\times10^{-6}$), (10.0,\ $10^{-7}$), (2.5,\ $10^{-8}$).		}
		\label{fig:O_water_OH}
	\end{figure*}
\begin{figure}
		\centering
		\includegraphics[width=0.495\textwidth]{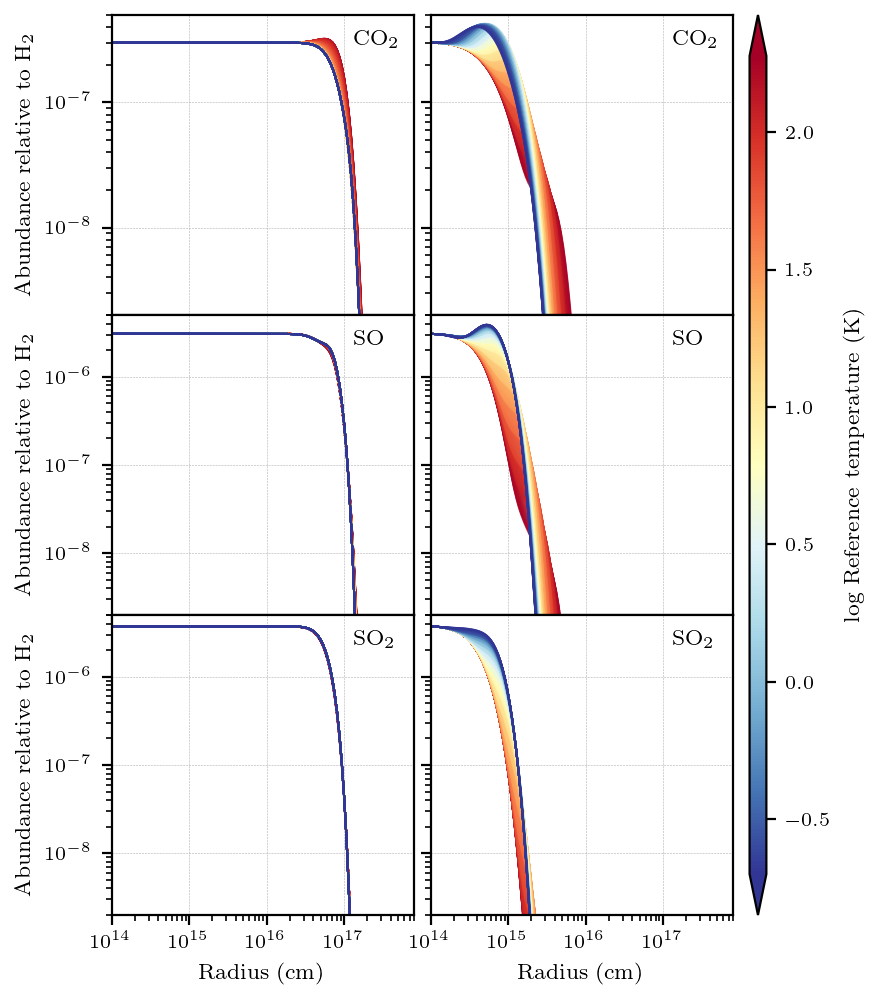}
		\caption{Fractional abundance profiles of CO$_2$, SO, and SO$_2$ in O-rich outflows, for a high and low outflow density: $(\vexp \ [{\rm \kms}],\Mdot \ [{\rm \Msolyr}])$: \textit{left} (25.0,\ $5\times10^{-5}$) and \textit{right} (2.5,\ $10^{-8}$).}
		\label{fig:O_CO2_SO_SO2}
	\end{figure}

	In O-rich AGB outflows, the chemistry is dominated by reactions with the parent species H$_2$O and its daughter OH. Fig.\ \ref{fig:O_water_OH} shows the abundance profiles of these species, arranged according to decreasing outflow density. For the highest density (left panels), the H$_2$O abundance drops drastically around $2\times 10^{17}\,{\rm cm}$ due to photodissociation into OH. Consequently, at the same radius, the OH abundance peaks in the outflow. When the outflow density is lower, H$_2$O is destroyed closer to the star. The location of the peak in the OH abundance shifts accordingly, following the location of H$_2$O photo-destruction. The governing temperature profile has a larger effect on the abundance profiles for outflows with low density. For the lowest density (right panels) the abundance of OH in the coolest model (blue curves) peaks at a radius of about $10^{15}$\,cm, while for warmer outflows (red curves) the peak lies about an order of magnitude further out in the outflow. This shift in location of the OH peak with temperature is consistent with the change in the extent of H$_2$O. 
	\\ \indent A similar trend is also visible in the abundance profiles of other parent-daughter pairs, e.g., for the pairs HCN-CN and NH$_3$-NH$_2$ (see Figs.\ in Supplementary Material). The abundance of the parents remains roughly its initial value until photodissociation, with a larger envelope size in warmer outflows at lower outflow density. At the location where the parents are destroyed, the daughter species are formed. However, the degree of temperature dependence of the abundance profile depends on the parent-daughter pair. For example, for the O-rich parent H$_2$S (Fig.\ \ref{fig:O_H2S_HS})  the temperature dependence is prominently less strong. We discuss this further in Sect.\ \ref{sect:temp_dependence}. 
	\\ \indent Not all parent species show this temperature dependence. Fig.\ \ref{fig:O_CO2_SO_SO2} shows the abundance profiles of parent species CO$_2$, SO, and SO$_2$ for a high and low outflow density. At high density (left panels), the abundance profiles do not significantly change with temperature, similar to H$_2$O. However, at low density (right panels), we now find that the abundances increase for cool outflows (blue curves), extending the envelopes size. This is caused by reactions with OH and its temperature-dependent abundance profile, as we elaborate on in Sect.\ \ref{sect:temp_dependence}. 
	\begin{figure*}
		\centering
		\includegraphics[width=0.95\textwidth]{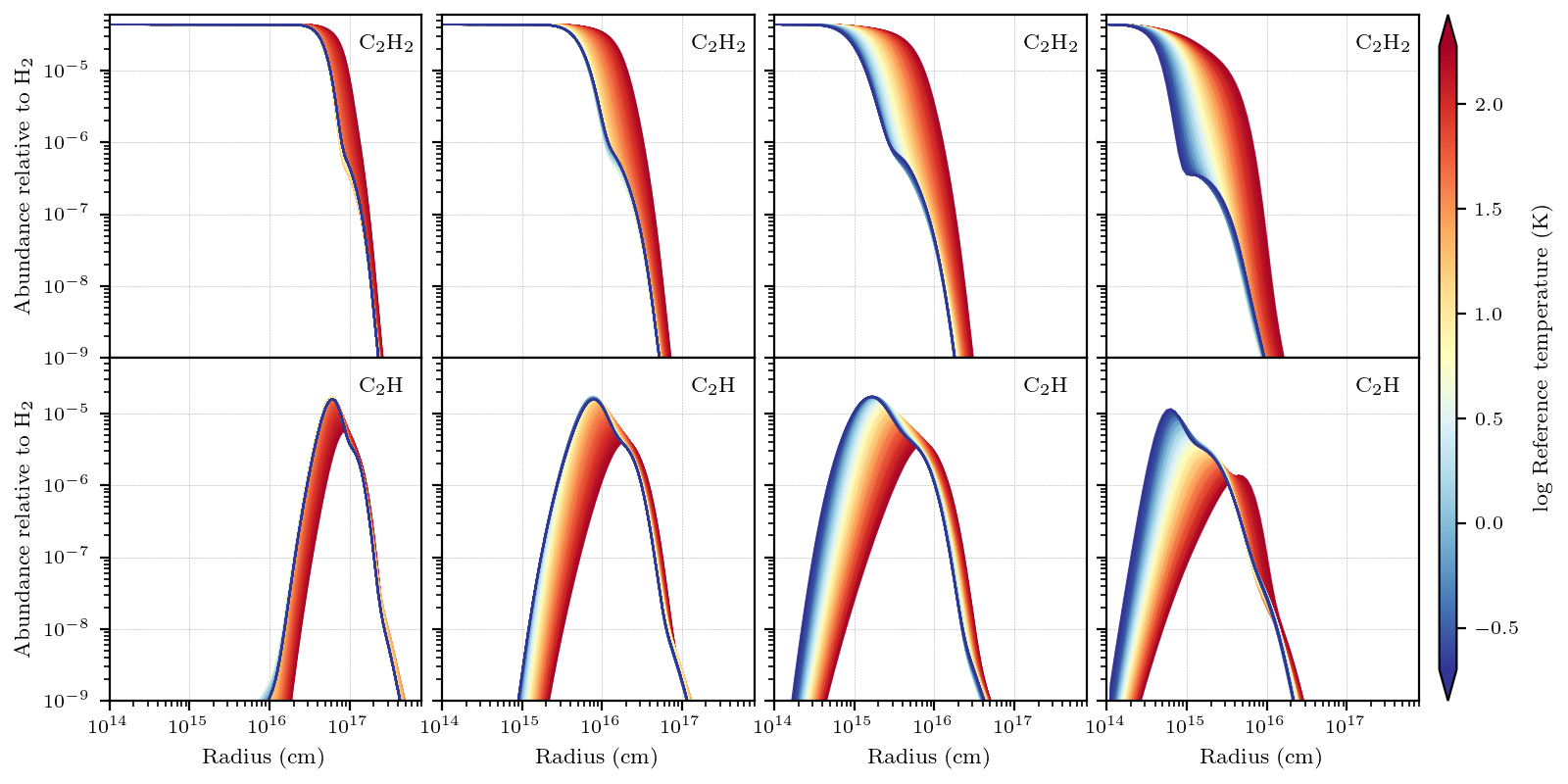}
		\caption{Fractional abundance profiles for C$_2$H$_2$ and C$_2$H in C-rich outflows, going from high to low outflow density. From \textit{left} to \textit{right} $(\vexp \ [{\rm \kms}],\Mdot \ [{\rm \Msolyr}])$: (25.0,\ $5\times10^{-5}$), (17.5,\ $2\times10^{-6}$), (10.0,\ $10^{-7}$), (2.5,\ $10^{-8}$).}
		\label{fig:C_C2H2_C2H}
	\end{figure*}
	\begin{figure*}
		\centering
		\includegraphics[width=0.95\textwidth]{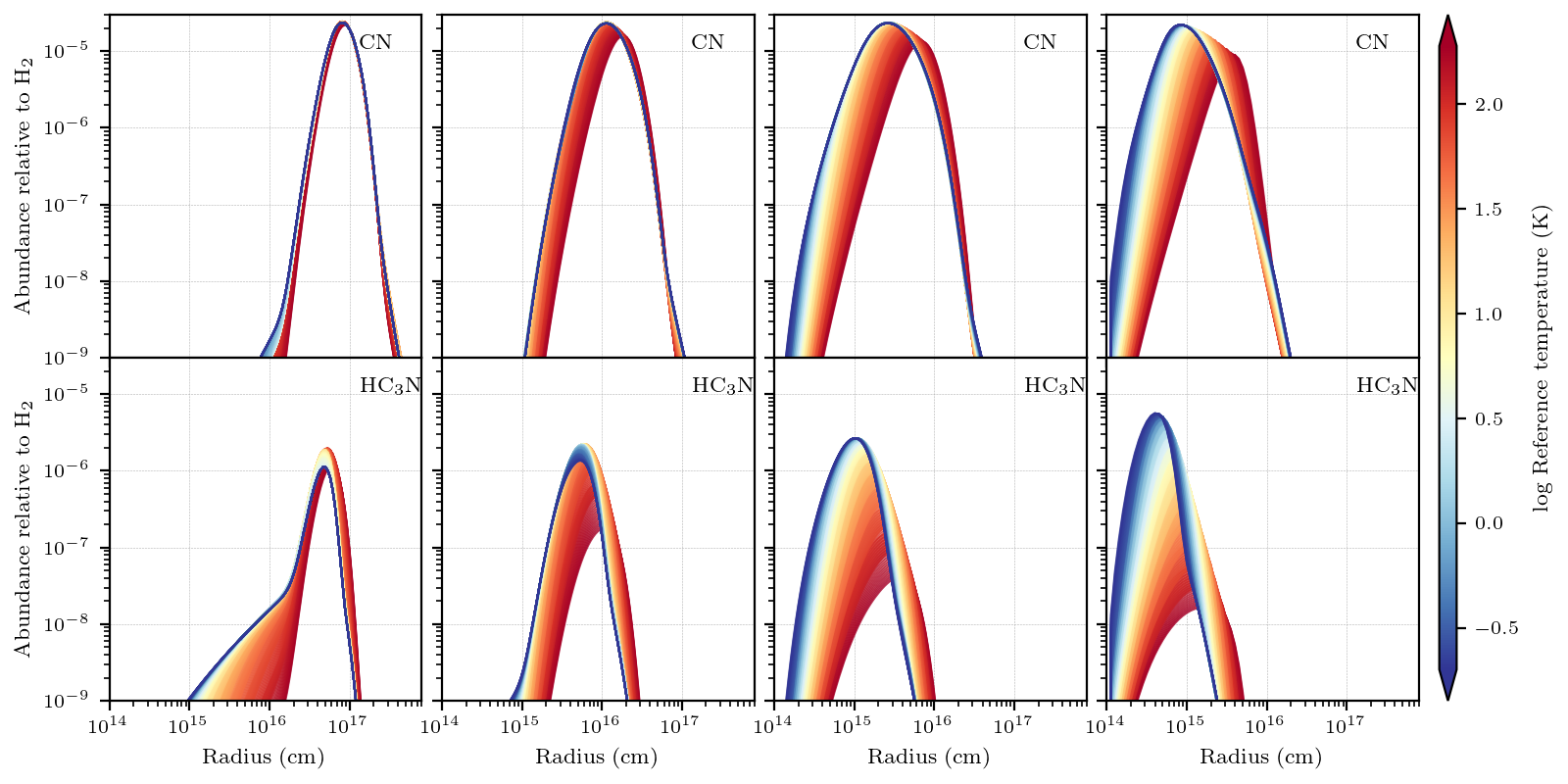}
		\caption{Fractional abundance profiles for CN and HC$_3$N in C-rich outflows, going from high to low outflow density. From \textit{left} to \textit{right} $(\vexp \ [{\rm \kms}],\Mdot \ [{\rm \Msolyr}])$: (25.0,\ $5\times10^{-5}$), (17.5,\ $2\times10^{-6}$), (10.0,\ $10^{-7}$), (2.5,\ $10^{-8}$). }
		\label{fig:C_cyanopolyynes}
	\end{figure*}

	\subsection{C-rich outflows}\label{sect:results_C}
	Chemistry in C-rich outflows is more diverse, since carbon is very reactive, readily producing a large variety of carbon-based molecules and ions. In Fig.\ \ref{fig:C_C2H2_C2H}, the abundance profiles of the parent species C$_2$H$_2$ and its daughter C$_2$H are shown, arranged according to decreasing outflow density. Similar to H$_2$O and OH in O-rich outflows, the abundance of the daughter C$_2$H peaks at the location where the parent C$_2$H$_2$ is photodissociated. For lower outflow densities, the parent species is photodissociated closer to the star, decreasing its envelope size and jointly shifting the peak in the abundance of the daughter species. The temperature profile again has a stronger effect on the shape of the abundance profile for low-density outflows (right panels of Fig.\ \ref{fig:C_C2H2_C2H}): for the warmer models (red curves) the decrease in abundance of C$_2$H$_2$ occurs further out in the wind, as compared to the cooler models (blue curves). Other parent-daughter pairs show a similar trend, such as HCN-CN, NH$_3$-NH$_2$ H$_2$O-OH, H$_2$S-HS, and CH$_4$-CH$_3$ (see Figs.\ in Supplementary Material).
	\\ \indent Fig.\ \ref{fig:C_cyanopolyynes} shows the abundance profiles of the species CN and HC$_3$N, arranged according to decreasing outflow density. Cyanopolyynes, of which HC$_3$N is the first in the sequence, are generally formed by adding CN, formed by the photodissociation of the parent HCN, to polyynes, starting off from the parent species C$_2$H$_2$ \citep{Agundez2017}: 
	\begin{flalign}
		&{\rm HCN} + {h\nu} \longrightarrow {\rm CN + H},& \label{eq:HCN}\\
		&{\rm C}_{2n}{\rm H_2} + {\rm CN} \longrightarrow {\rm HC}_{2n+1}{\rm N+H},& \label{eq:cyanopolyynes}
	\end{flalign}
	This formation mechanism makes them $n^{\rm th}$-generation daughter species. In Fig.\ \ref{fig:C_cyanopolyynes}, we find that the abundance profile of HC$_3$N at different densities shows the same general behaviour as CN, since the temperature dependence of its abundance profile is inherited from CN. 

	\section{Discussion} \label{sect:discussion}
	In this Section, we explain the different trends found in the abundance profiles from Sect.\ \ref{sect:results}, i.e., analysing the dependence on outflow density and temperature. Moreover, we calculate the molecular envelope size of the parent species and compare these to the literature. Finally, we discuss the effect of the outflow temperature and density on the possible detectability of certain species. We note that the individual abundances of species heavily depend on the set of parent species and the other assumptions made in the model (Sect.\ \ref{sect:chem_model}).
	
	\subsection{Variation in abundance profiles}\label{sect:var_in_abs}
	We find that abundance profiles depend on the outflow density due to the extinction-dependent photodissociation rate, which is proportional to the density. Moreover, abundance profiles can be temperature dependent {due to the presence of an energy barrier in the main reaction channel. These temperature dependences can be inherited by subsequent generations of daughter species, consequently becoming temperature dependent themselves.}

	\subsubsection{Dependence on density}\label{sect:dens_dependence}
	The variations in abundance profiles for different outflow densities is regulated by photodissociation, since its rate depends on the extinction in the outflow, proportional to the density. Hence, when the density is high, external UV photons experience a larger extinction, lowering the photodissociation rate. This means that in higher density outflows, species will be photodestroyed further away from the star, compared to low density outflows, resulting in larger envelope extents for the parent species. As a direct consequence, the abundances of the daughter species peak further out in the outflow as well.
	\\ \indent For lower density outflows, the abundance profiles depend more strongly on temperature. In these outflows, photodissociation occurs closer to the star, where the temperature is still higher (Eq.\ \ref{eq:temp_profile}), as is generally also the case for reaction rates. Hence, this increases the diversity in chemical pathways.
	
	\subsubsection{Dependence on temperature}\label{sect:temp_dependence}
	For models with the same outflow density, the variation in the shape of the abundance profiles is caused by the effect of a different temperature profile throughout the outflow, since photodissociation will occur at the same distance from the star. The temperature profile is primarily set by the exponent, $\eps$, rather than the stellar temperature, $\Tstar$, the latter mainly influencing the temperature in the inner wind (Eq.\ \ref{eq:temp_profile}). Moreover, the effect of changing the temperature profile is largest for models with a lower outflow density. Therefore, we focus in this Section on the effect of $\eps$ in a low-density outflow ($\vexp = 2.5\,{\rm \kms}$, $\Mdot = 10^{-8}\,{\rm \Msolyr}$). For higher density outflows, the same reasoning holds, only the effect is less strong.
	\\ \indent {We find that the temperature dependence of the abundance profile of certain species} results from reactions involving energy barriers ($\gamma \neq 0$ in Eq.\ \ref{eq:rate_coeff}). 
	For temperature profiles with a lower value of $\eps$, the outflow stays warmer throughout compared to high $\eps$ values (see Eq.\ (\ref{eq:temp_profile}) and, e.g., right panel of Fig.\ \ref{fig:grid}). Hence, energy barriers for certain reactions can be overcome more easily and in a larger fraction of the outflow. This results in an abundance increase or decrease for certain species in the warmer outflows, depending on which reaction in the pathway contains the energy barrier. 
	\\ \indent In the case of H$_2$O (Fig.\ \ref{fig:O_water_OH}, right panel), the abundance increases in the inner part of the simulated region for models with a low $\eps$ value (warmer outflows, red curves), compared to models with a high $\eps$ (cooler outflows, blue curves). This is linked to the consecutive hydrogenation of oxygen to form water, where large energy barriers need to be overcome:
	\begin{flalign}
	&{\rm O  + H_2} \rightarrow {\rm OH + H} \quad \quad \ \ (\gamma = 3150\,{\rm K}),& \label{eq:OH}\\ 
	&{\rm OH + H_2} \rightarrow {\rm H_2O + H}\quad \quad (\gamma = 1736\,{\rm K}).&\label{eq:H2O}
	\end{flalign}
	Thus, when the temperature in the outflow is still high enough after photodissociation of H$_2$O into OH, OH can be converted back to H$_2$O (reaction \ref{eq:H2O}). Consequently, the envelope size of H$_2$O in cool outflows is mainly set by photodissocation and is therefore smaller than in warmer outflows. The abundance profile of OH is linked to this, so that for warm outflows, its peak in abundance is located further from the star. 
	\\ \indent A similar explanation holds for the abundance profile of C-rich parent species C$_2$H$_2$ and its daughter C$_2$H (right panels of Fig.\ \ref{fig:C_C2H2_C2H}). The main reaction is:
	\begin{equation}\label{eq:C2H2}
	{\rm C_2H} + {\rm H_2} \rightarrow {\rm C_2H_2} + {\rm H}  \quad \quad \ \ (\gamma = 130\,{\rm K}).
	\end{equation}
	Again, C$_2$H can be converted back to C$_2$H$_2$ when the temperature is high enough at that location of C$_2$H$_2$ photodissociation. Additionally, in Fig.\ \ref{fig:C_C2H2_C2H}, we identify for the cooler models (blue curves) a kink in the abundance profile for both C$_2$H$_2$ and C$_2$H at a radius of $\sim 10^{15}\,{\rm cm}$. This is caused by reactions involving C$_2$H$_3^+$ producing C$_2$H$_2$, and ${\rm HC_3N}$ producing C$_2$H:
	\begin{flalign}
	&{\rm C_2H_3^+} + {\rm e^-} \rightarrow {\rm C_2H_2 + H} \quad \quad \ \ (\beta = -0.84),& \label{eq:C2H3+}\\ 
	&{\rm HC_3N} + h\nu \rightarrow {\rm C_2H + CN}, &\label{eq:HC3N_hnu}
	\end{flalign}
	together with the interplay between C$_2$H$_2$ and C$_2$H (reaction \ref{eq:C2H2}).
	Reactions (\ref{eq:C2H3+}) and (\ref{eq:HC3N_hnu}) become important around a radius of $\sim 10^{15}\,{\rm cm}$ only in cooler outflows, operating as an additional source of C$_2$H$_2$ and C$_2$H molecules, increasing their abundances.	
	\\ \indent Hydrogenation reactions underpin behaviour in more parent-daughter pairs, where the parent can be reformed by hydrogenation of its daughter: for both chemical types this includes HCN-CN, H$_2$S-HS, and NH$_3$-NH$_2$-NH. For C-rich chemistry, this also holds for CH$_4$ and its daughters (see Figs.\ in Supplementary Material). However, the abundance profile of the H$_2$S-HS pair in O-rich outflows is significantly less temperature dependent compared to the other parent species (see Fig.\ \ref{fig:O_H2S_HS}). This is caused by the species H$_2$S$^+$ and H$_3$S$^+$, connected via the following reaction, containing an energy barrier:
	\begin{equation}
	{\rm H_2S^+} + {\rm H_2} \rightarrow {\rm H_3S^+} + {\rm H}  \quad \quad \ \ (\gamma = 2900\,{\rm K}).
	\end{equation}
	Hence, H$_2$S$^+$ and H$_3$S$^+$ will depend inversely on temperature (see Fig.\ \ref{fig:O_H2S+_H3S+}). Adding an electron to both H$_2$S$^+$ and H$_3$S$^+$ results in the formation of H$_2$S, hence largely removing the temperature dependence.
	\\ \indent {The variation in the abundance profile due to energy barriers} can be inherited by the subsequent generations of daughter species. For example, the temperature dependence of H$_2$O and OH regulates the abundance profiles of parents species in O-rich outflows, such as CO$_2$, CS, SO, SO$_2$, and SiO, three of which are shown in Fig.\ \ref{fig:O_CO2_SO_SO2}. The higher abundance of CO$_2$ around $r=10^{15}\,$cm in cool outflows is due to the reaction 
	\begin{equation}\label{eq:CO2}
	{\rm CO} + {\rm OH} \rightarrow {\rm CO_2} + {\rm H},
	\end{equation}
	since at this radius, also the abundance of OH is higher for the cooler models. The situation is similar for SO and SO$_2$ (Fig.\ \ref{fig:O_CO2_SO_SO2}) , as they are formed by the following reactions:
	\begin{flalign}
	&{\rm S  + OH} \rightarrow {\rm SO + H}& \label{eq:SO}\\ 
	&{\rm SO + OH} \rightarrow {\rm SO_2 + H}& \label{eq:SO2}.
	\end{flalign}
	Although their reaction rates are temperature independent ($\beta=\gamma=0$ in Eq.\ \ref{eq:temp_profile}), the large availability of OH imposes a temperature dependence on the abundance profiles of SO and SO$_2$. We note that for SO, \cite{Danilovich2016} have observed an increase in abundance in the outer part of the envelope in sources with higher-density outflow. We do not find such an increase here, which could be due to our models including gas-phase chemistry in a smooth outflow only. Models that include dust-gas chemistry or a clumpy outflow can reproduce this behaviour \citep{VandeSande2019,Danilovich2020}. However, including these mechanisms is beyond the scope of this work. In the case of the parent species CS (Fig.\ \ref{fig:O_CS_OCS}), the dependence on OH is related to destruction rather than production. CS can react with OH, forming OCS. Since OH is more abundant for cooler outflows at radii $\leq 10^{15}\,{\rm cm}$, CS will be more rapidly destroyed. In warmer outflows, CS will not be as readily destroyed by OH and will maintain its abundance until photodissociation. In Sect.\ \ref{sect:detectability} we elaborate on the detectability of OCS.
	\\ \indent In the C-rich outflows, an analogous situation is found for the cyanopolyynes. The abundance profiles of the cyanopolyyne ${\rm HC_3N}$ together with CN are shown in Fig.\ \ref{fig:C_cyanopolyynes} and since ${\rm HC_3N}$ is formed via reaction (\ref{eq:cyanopolyynes}), it inherits the profile of CN. 
	\\ \indent {We note that reaction rates can be temperature dependent via the parameter $\beta$ in Eq.\ (\ref{eq:temp_profile}). A positive value will make the reaction more likely to proceed at higher temperatures, as a negative value will have the opposite effect. However, we found that this effect is negligible compared to the impact of energy barriers. }

	\subsection{Envelope size of parent species}\label{sect:env}
	\begin{table*}
		\begin{center}
			\caption{Slope $a$ and intercept $b$ for the linear fit to the sizes of parents' molecular envelope extents (Eq.\ \ref{eq:loglin_fit}).}
			\begin{tabular}{   l  c  c  c  l c  c     }
				\hline \hline \\[-2ex]
				Carbon-rich &  & & & Oxygen-rich & & \\ 
				{Parent species} & {$a$} & {$b$} & &	{Parent species} & {$a$} & {$b$}  \\ \hline
				CO & $0.5096 \pm0.0004 $ & $ 20.641\pm0.003$ & &CO & $0.5222 \pm0.0003 $ & $ 20.545\pm0.002$ \\ 
				C$_2$H$_2$ & $0.507 \pm0.002 $ & $ 19.42\pm0.01$ & &H$_2$O & $0.534 \pm0.001 $ & $ 19.78\pm0.01$ \\ 
				HCN & $0.528 \pm0.002 $ & $ 19.64\pm0.01$ & &N$_2$ & $0.48 \pm0.001 $ & $ 19.79\pm0.01$ \\ 
				N$_2$ & $0.439 \pm0.002 $ & $ 19.52\pm0.01$ & &SiO & $0.619 \pm0.001 $ & $ 20.43\pm0.01$ \\ 
				SiC$_2$ & $0.448 \pm0.002 $ & $ 19.41\pm0.01$ & &H$_2$S & $0.668 \pm0.001 $ & $ 20.42\pm0.01$ \\ 
				CS & $0.483 \pm0.001 $ & $ 19.62\pm0.01$ & &SO$_2$ & $0.641 \pm0.001 $ & $ 20.37\pm0.01$ \\ 
				SiS & $0.443 \pm0.002 $ & $ 19.27\pm0.01$ & &SO & $0.632 \pm0.001 $ & $ 20.44\pm0.01$ \\ 
				SiO & $0.586 \pm0.001 $ & $ 20.0\pm0.01$ & &SiS & $0.511 \pm0.002 $ & $ 19.64\pm0.01$ \\ 
				CH$_4$ & $0.584 \pm0.001 $ & $ 20.02\pm0.01$ & &NH$_3$ & $0.612 \pm0.001 $ & $ 20.15\pm0.01$ \\ 
				H$_2$O & $0.489 \pm0.002 $ & $ 19.47\pm0.01$ & &CO$_2$ & $0.56 \pm0.001 $ & $ 20.03\pm0.01$ \\ 
				C$_2$H$_4$ & $0.63 \pm0.001 $ & $ 20.17\pm0.01$ & &HCN & $0.56 \pm0.001 $ & $ 19.88\pm0.01$ \\ 
				NH$_3$ & $0.545 \pm0.001 $ & $ 19.66\pm0.01$ & &CS & $0.461 \pm0.003 $ & $ 19.24\pm0.02$ \\ 
				C$_2$H$_4$ & $0.63 \pm0.001 $ & $ 20.17\pm0.01$ & & & &  \\   \hline
			\end{tabular}
			\label{tab:fit_env_ext_all}
		\end{center}
	\end{table*}
	Since parent species are destroyed by photodissociation in the outer part of the outflows, the size of their resulting molecular envelope depends on the density of the outflow. Moreover, due to the hydrogenation reactions with daughter species, parent species' envelopes can also be dependent on the temperature profile.
	\\ \indent 	Generally, the size of the envelope is defined as the radius at which the initial abundance $f_0$ has dropped to a value $f_0/e$, referred to as the $e$-folding radius, $R_e$, which is similar to a scale height. We calculated $R_e$ for all models and analysed the results as a function of $\Mdot/\vexp$, a measure for density (Eq.\ \ref{eq:density}). {The envelope sizes are set by photodissociation, and hence they depends on the interstellar radiation field and its extinction in the outflow. We elaborate on the extinction of ISM UV radiation in Appendix \ref{sect:photodiss}.}
	
	\subsubsection{General trends}\label{sect:env_general_trends}
\begin{figure}
		\centering
		\includegraphics[width=0.49\textwidth]{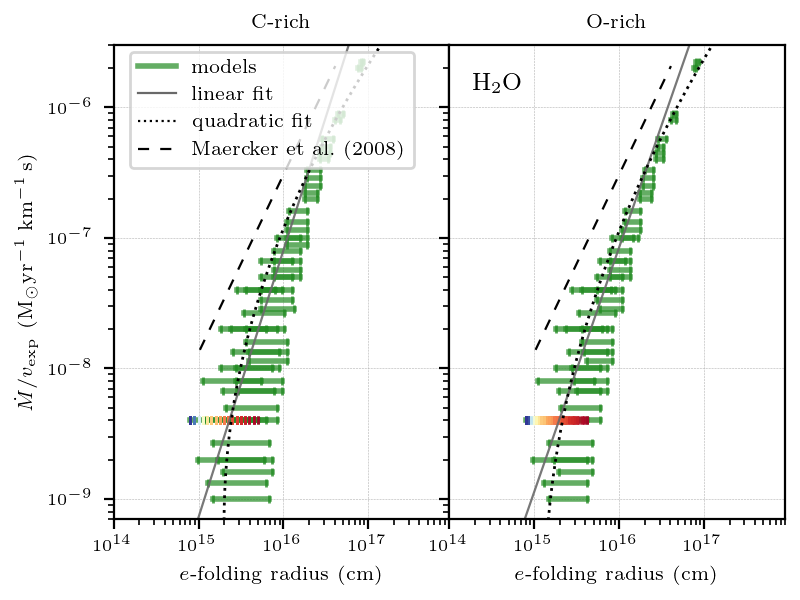}
		\caption{$e$-folding radii of H$_2$O for the C-rich (\textit{left}) and O-rich (\textit{right}) models in green. For one density the reference temperature (see Sect.\ \ref{sect:results}) is indicated with the colour bar; the blue side indicates the cold outflows ($\eps=1.0$) and the red side the warm outflows ($\eps = 0.3$). The full grey line represents the linear fit to the data (Eq.\ \ref{eq:loglin_fit}, Table \ref{tab:fit_env_ext_all}), the dotted line represents the linear fit (Eq.\ \ref{eq:logq_fit}, Table \ref{tab:fit_quadr_env_ext_all}). The black dashed line follows Eq.\ (\ref{eq:efolding_H2O}) {for the density range of the source used by \protect\cite{Maercker2008}}.}
		\label{fig:H2O_env}
	\end{figure}

	Overall, due to the dependence of photodissociation on density, we find a roughly linear dependence of $\log \, R_e$ on $\log(\Mdot/\vexp)$ for the parent species of C-rich as well as O-rich outflows, where a higher outflow density results in a larger envelope extent. Hence, we can fit the linear relation to the envelope sizes:
	\begin{equation}\label{eq:loglin_fit}
		\log  R_e= a\log \frac{\Mdot}{\vexp}+b,
	\end{equation}
	with fitting parameters $a$ and $b$, representing the slope and intercept, respectively. We used a linear regression routine ({\footnotesize \verb|scipy.stats.linregress|}, \citealp{scipy2020S}) to find the fitting parameters and their standard deviation for the envelope sizes of all parent species. The results can be found in Table \ref{tab:fit_env_ext_all}. 
	\begin{figure}
		\centering
		\includegraphics[width=0.49\textwidth]{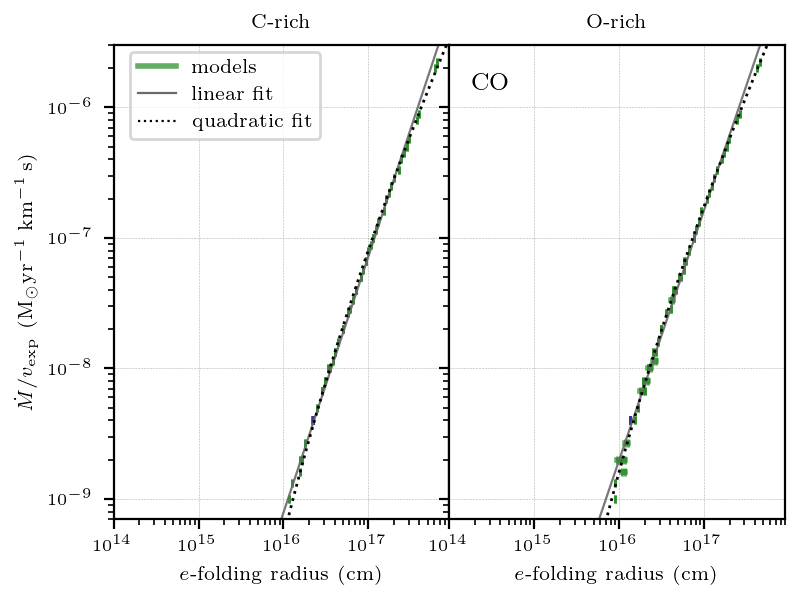}
		\caption{$e$-folding radii of CO for the C-rich (\textit{left}) and O-rich (\textit{right}) models in green. The full grey line represents the linear fit to the data (Eq.\ \ref{eq:loglin_fit}, Table \ref{tab:fit_env_ext_all}), the dotted line represents the linear fit (Eq.\ \ref{eq:logq_fit}, Table \ref{tab:fit_quadr_env_ext_all}).}
		\label{fig:CO_env}
	\end{figure}
	\\ \indent In Fig.\ \ref{fig:CO_env}, the envelope size of the parent species CO as a function of the density measure $\Mdot/\vexp$ is shown, for both C-rich and O-rich outflows, together with the fit to the data. We see that the relation between the $e$-folding radius and density is nearly perfectly linear. This is because for CO the size of the envelope is predominantly influenced by photodissociation, independent of temperature, because the molecular bond in CO is strong. This makes that other reactions are not able to significantly change the amount of the overly abundant CO, and hence the envelope size.
	\\ \indent The envelope size of, e.g., H$_2$O shown in Fig.\ \ref{fig:H2O_env}, diverges from the linear relation, especially for the models with lower outflow density. For these models, larger values of $R_e$ are found, because the abundance profiles of the specific parent species depends on the temperature profile, as explained in Sect.\ \ref{sect:temp_dependence}. Hence, their envelope sizes do as well. In the case of H$_2$O, this is due to the energy barriers in its formation pathway (reactions \ref{eq:OH} and \ref{eq:H2O}). The same reasoning holds for most parents, such as  NH$_3$, H$_2$S, and HCN for both wind types, CH$_4$ and C$_2$H$_2$ in C-rich winds, and CO$_2$ in O-rich winds. For Si-bearing parent species the temperature dependence of the envelope sizes is indirectly related to energy barriers, since they interact with species such as H$_2$O, OH, HCN, and C$_2$H$_2$ (Sect.\ \ref{sect:temp_dependence}). 
	\\ \indent Consequently, we find that the quadratic relation
	\begin{equation}\label{eq:logq_fit}
		\log  R_e= \gamma\left(\log \frac{\Mdot}{\vexp}\right)^2+\alpha\log \frac{\Mdot}{\vexp}+\beta
	\end{equation}
	better fits the envelope sizes. The resulting fitting parameters ($\gamma$, $\alpha$, and $\beta$) per parent species can be found in Table \ref{tab:fit_quadr_env_ext_all}. When these fitting results are compared to the results of the linear fit in Table \ref{tab:fit_env_ext_all}, respectively, we find that the parameters $\alpha$ and $a$ are of the same order of magnitude, as are parameters $\beta$ and $b$, from which we conclude that the linear approach (Eq.\ \ref{eq:loglin_fit}) is reasonable first approximation.
	\begin{table*}
		\begin{center}
			\caption{Fitting parameters of the quadratic fit to the sizes of parents' molecular envelope extents (Eq.\ \ref{eq:logq_fit}).}
			\begin{tabular}{   l  c  c  c c l c  c  c   }
				\hline \hline \\[-2ex]
				Carbon-rich & & & & & Oxygen-rich & & & \\ 
				{Parent species} & {$\gamma$} & {$\alpha$} & $\beta$ &  &	{Parent species} & {$\gamma$} & {$\alpha$} & $\beta$ \\ \hline
				CO & 1.02 & 0.03 & 22.5 & & CO & 1.01 & 0.03 & 22.32 \\ 
				C$_2$H$_2$ & 2.45 & 0.13 & 26.49 & & H$_2$O & 2.12 & 0.11 & 25.5 \\ 
				HCN & 2.38 & 0.12 & 26.38 & & N$_2$ & 2.13 & 0.11 & 25.74 \\ 
				N$_2$ & 2.46 & 0.14 & 26.89 & & SiO & 1.44 & 0.06 & 23.39 \\ 
				SiC$_2$ & 2.46 & 0.14 & 26.75 & & H$_2$S & 1.89 & 0.08 & 24.83 \\ 
				CS & 2.34 & 0.13 & 26.4 & & SO$_2$ & 1.64 & 0.07 & 23.97 \\ 
				SiS & 3.0 & 0.17 & 28.62 & & SO & 1.5 & 0.06 & 23.56 \\ 
				SiO & 2.37 & 0.12 & 26.52 & & SiS & 2.77 & 0.15 & 27.82 \\ 
				CH$_4$ & 2.45 & 0.13 & 26.81 & & NH$_3$ & 2.06 & 0.1 & 25.37 \\ 
				H$_2$O & 2.53 & 0.14 & 26.93 & & CO$_2$ & 1.76 & 0.08 & 24.37 \\ 
				C$_2$H$_4$ & 2.31 & 0.11 & 26.3 & & HCN & 2.06 & 0.1 & 25.29 \\ 
				NH$_3$ & 2.43 & 0.13 & 26.54 & & CS & 3.25 & 0.19 & 29.3 \\ 
				C$_2$H$_4$ & 2.31 & 0.11 & 26.3 & & & & &  \\  \hline
			\end{tabular}
			\label{tab:fit_quadr_env_ext_all}
		\end{center}
	\end{table*}

	\subsubsection{Effect of CO self-shielding} \label{sect:COselfshielding}
	\begin{figure}
		\centering
		\includegraphics[width=0.49\textwidth]{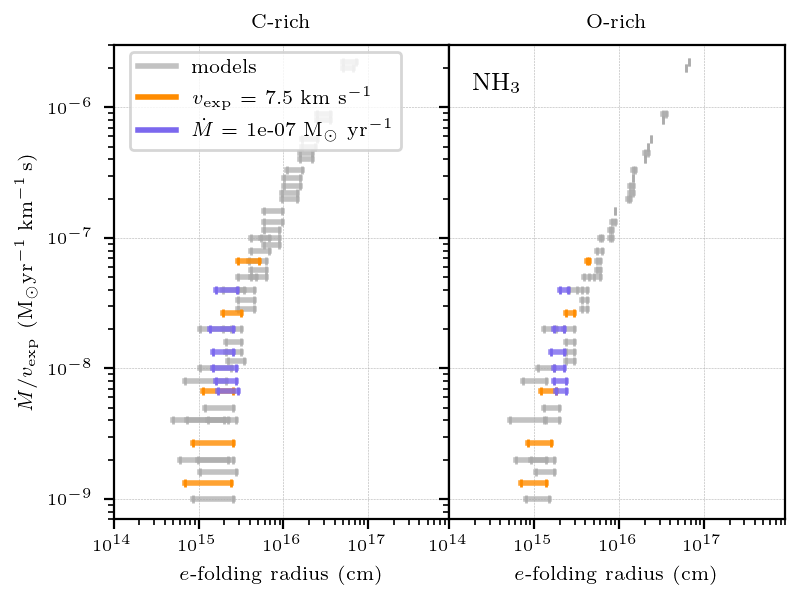}
		\caption{$e$-folding radii of NH$_3$ for the C-rich (\textit{left}) and O-rich (\textit{right}) models. In orange, models with an expansion velocity of 7.5\,${\rm \kms}$ are highlighter, in purple, models with a moss-loss rate of $10^{-7}\,{\rm \Msolyr}$.} 
		\label{fig:NH3_env}
	\end{figure}
	The spread on the $e$-folding radii of the parent species, and deviation from the linear relation, is also partially caused by CO self-shielding. CO photodissociation is dominated by line absorptions and can only be photodissociated at specific wavelengths. Hence, thanks to the the high CO abundance in AGB outflows, CO shields itself from the incoming radiation, leading to so-called self-shielding. As mentioned in Sect.\ \ref{sect:chem_model}, the CO self-shielding is implemented in the model according to the single-band approximation from \cite{MJ1983}, meaning we take into account the line at $1000\,\AA$. The photodissociation rate of CO is velocity dependent, due to the Doppler shift of the moving medium. Hence, the photodissociation rate is higher for larger expansion velocities \citep{MJ1983}.
	\\ \indent  In Fig.\ \ref{fig:NH3_env}, the $e$-folding radii are given for the parent NH$_3$, with specific models highlighted: in orange models with an expansion velocity of 7.5\,$\kms$, in purple with a mass-loss rate of $10^{-7}\,\Msolyr$. Hence, in both cases the highlighted models have a different density, while $\Mdot$ and $\vexp$, respectively, are kept constant. For the models with constant expansion velocity, the general trend is retrieved, where higher densities result in larger envelopes. However, for the models with constant mass-loss rate, the opposite is found. At constant mass-loss rate, the photodissociation rate increases with increasing expansion velocity (decreasing density, see Eq.\ \ref{eq:density}), due to the velocity-dependent CO self-shieling. This causes CO to be destroyed closer to the star. An earlier photodissociation of CO in the outflow produces a larger amount of C and O atoms closer to the star. These atoms are reactive, driving the chemistry including the formation of parent species. As a result, the envelopes size of other parent species are larger for higher velocities when the mass-loss rate is kept constant, despite the species being photodissociated closer to the star. This effect is again stronger at low density (Sect.\ \ref{sect:dens_dependence}) , making $\log\, R_e$ diverge from the linear trend. 
	\\ \indent 
		\begin{figure}
		\centering
		\includegraphics[width=0.49\textwidth]{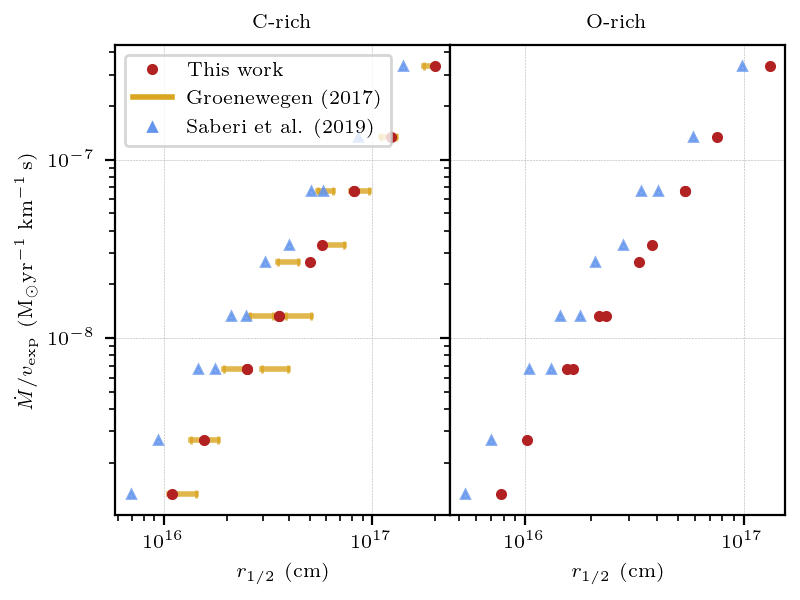}
		\caption{CO envelope size ($r_{1/2}$) found in this work (red circles), according to \protect\cite{Groenewegen2017} (yellow bars, left $T_{\rm ex} = 100\,{\rm K}$, right $T_{\rm ex} = 5\,{\rm K}$), and according to \protect\cite{Saberi2019} (blue triangle) for 12 overlapping models, for the C-rich (\textit{left}) and O-rich (\textit{right}) models.}
		\label{fig:CO_saberi}
	\end{figure} 
	In Fig.\ \ref{fig:CO_saberi}, we compare the single-band approximation for CO self-shielding used here to two previous studies: 
	\citeauthor{Groenewegen2017} (\citeyear{Groenewegen2017}, hereafter G17) used tabulated shielding functions of more CO photodissociation lines in their calculation, and therefore used different excitation temperatures, $T_{\rm ex}$ (5, 10, 50, 100\,K). \citeauthor{Saberi2019} (\citeyear{Saberi2019}, hereafter SVDB19) included mutual shielding by other species, next to multiple CO photodissociation lines, and assumed the excitation temperature to be equal to the gas kinetic temperature. Our grid contains 12 models with overlapping outflow density with both G17 and SVDB19, indicated in green in Fig.\ \ref{fig:grid}. For the study of G17, only an overlapping initial CO abundance was found with our C-rich models, contrary to SVDB19. We could not compare our results with the pioneering work of \cite{Mamon1988}, because the input parameter spaces do not match. G17 and SVDB19 both use $r_{1/2}$ instead of $R_e$, which is the radius where the abundance has dropped half of the initial value. Fig.\ \ref{fig:CO_saberi} shows the $r_{1/2}$ of the 12 overlapping models, together with the values found by G17 and SVDB19. The yellow ranges indicate the results for different excitation temperatures of G17. The envelope sizes found in this work are slightly larger than those of SVDB19, {by a factor $1.5\pm0.1$ for the C-rich and $1.4\pm0.1$} for the O-rich models, on average. 
	The CO envelope sizes of G17 correspond quite well with the results found in this study, with a difference of about{ a factor of $1.0\pm0.2$}. Hence, we conclude that the single-band approximation is sufficiently accurate for a study of a gas-phase chemistry in CSEs, in spite of using the single band approximation, significantly reducing the computation time. 

	\section{Sensitivity analysis}\label{sect:sens}
	\begin{figure*}
		\centering
		\includegraphics[width=0.95\textwidth]{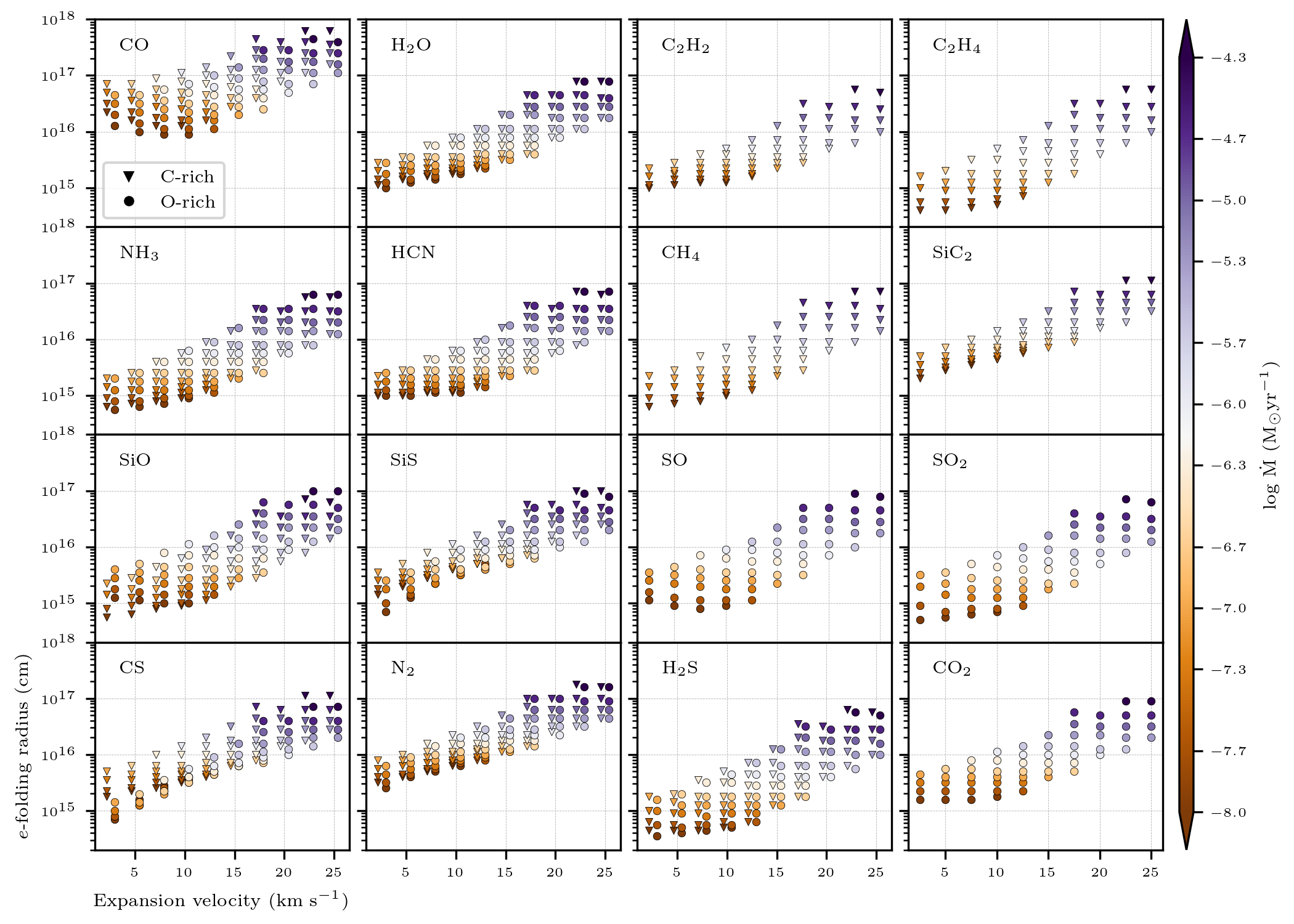}
		\caption{{Variation of the $e$-folding radii of parent species for different C-rich (triangles) and O-rich (circles) models. The temperature profile is fixed at $\Tstar = 2500\,$K and $\eps = 0.6$ (Eq.\ \ref{eq:temp_profile}). The mass-loss rate, indicated by the colour, and the expansion velocity can only take values from the discrete grid (see Fig.\ \ref{fig:grid} and Table \ref{tab:grid}). For better visibility and differentiation between C-rich and O-rich results, the symbols are centred around the corresponding expansion velocity for certain parents.}}
		\label{fig:sens_parents}
	\end{figure*}
	\begin{figure*}
		\centering
		\includegraphics[width=0.7\textwidth]{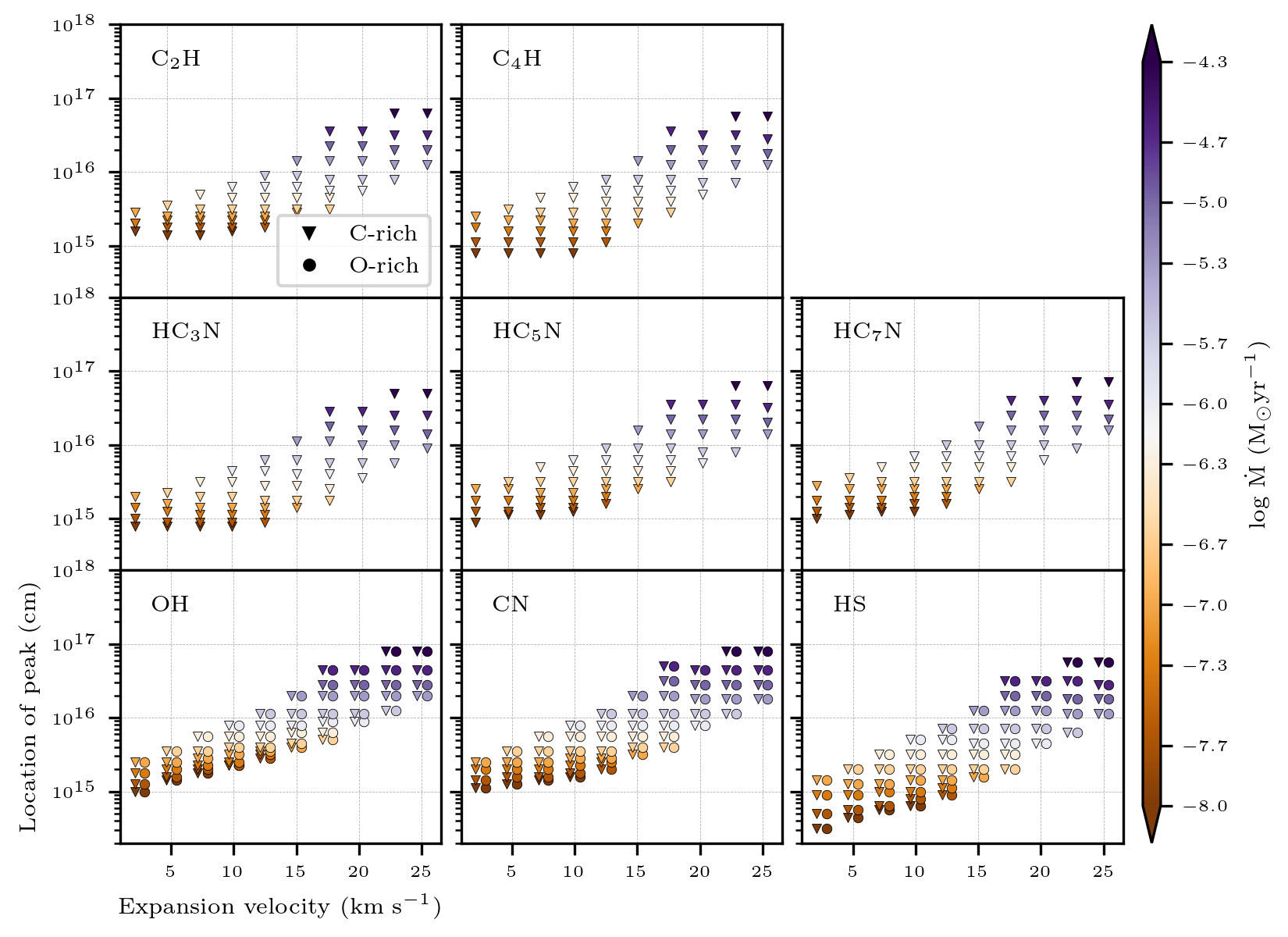}
		\caption{{Location of the peak in abundance profile of daughter species for different C-rich (triangles) and O-rich (circles) models, The temperature profile is fixed at $\Tstar = 2500\,$K and $\eps = 0.6$ (Eq.\ \ref{eq:temp_profile}). The mass-loss rate, indicated by the colour, and the expansion velocity can only take values from the discrete grid (see Fig.\ \ref{fig:grid} and Table \ref{tab:grid}). For better visibility and differentiation between C-rich and O-rich results, the symbols are centred around the corresponding expansion velocity for certain daughters. }}
		\label{fig:sens_daughters_r}
	\end{figure*}
	\begin{figure*}
		\centering
		\includegraphics[width=0.7\textwidth]{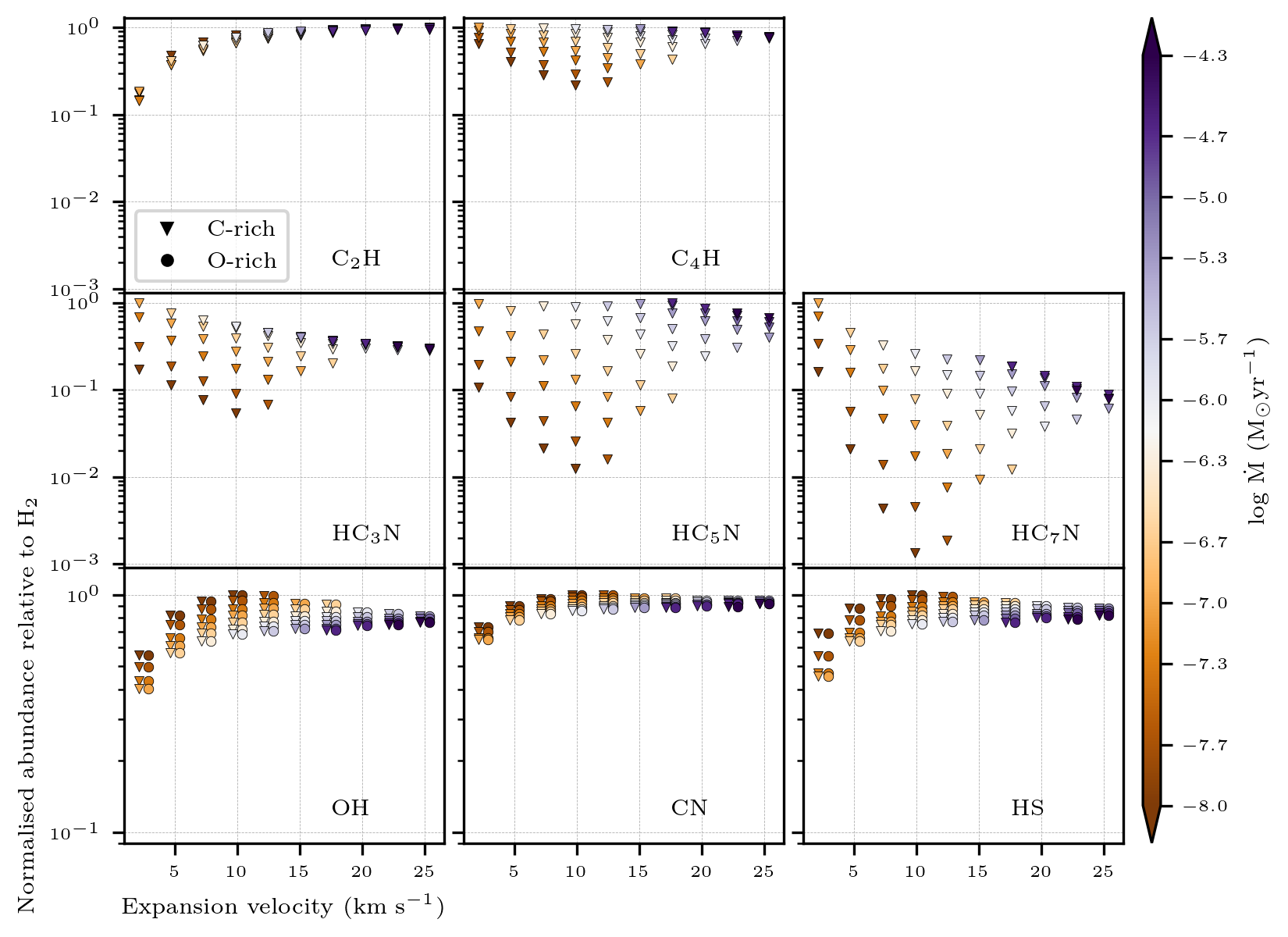}
		\caption{Peak abundance relative to H$_2$, normalised to maximum abundance in the panels, of daughter species for different C-rich (triangles) and O-rich (circles) models. The temperature profile is fixed at $\Tstar = 2500\,$K and $\eps = 0.6$ (Eq.\ \ref{eq:temp_profile}). The mass-loss rate, indicated by the colour, and the expansions velocity can only take values from the discrete grid (see Fig.\ \ref{fig:grid} and Table \ref{tab:grid}). For better visibility and differentiation between C-rich and O-rich results, the symbols are centred around the corresponding expansion velocity for certain daughters. Maximum abundances relative to H$_2$: C$_2$H: $8.846\times10^{-6}$, C$_4$H: $6.436\times10^{-7}$, HC$_3$N: $3.093\times10^{-6}$, HC$_5$N: $1.804\times10^{-7}$, HC$_7$N: $3.593\times10^{-7}$, OH: $4.974\times10^{-5}$, CN: $7.329\times10^{-8}$, HS: $2.571\times10^{-6}$.}
		\label{fig:sens_daughters_fracs}
	\end{figure*}
	{\noindent The mass-loss rate, expansion velocity, and temperature of AGB outflows are constrained from observations. Consequently, these values come with a certain uncertainty. For the expansion velocity, this uncertainty is quite small, since it can be determined accurately from spectral line widths. Mass-loss rates are generally estimated from the CO line emission in combination with radiative transfer modelling assuming spherical symmetry. This approach introduces large uncertainties, however the exact value is still under debate. One finds uncertainties from about a factor of 2, sometimes up to an order of magnitude (e.g., \citealp{KnappMorris1985,Ramstedt2008}). Since the mass-loss rate sets the outflow density and density in turn influences the photodissociation rate of chemical species (Sect.\ \ref{sect:dens_dependence}), uncertainties on the observationally estimated mass-loss rates will introduce uncertainties on the observed abundances. Retrieving the temperature profile of the outflow is done in a similar way as determining the mass-loss rate. The power law from Eq.\ (\ref{eq:temp_profile}) is assumed and fitted using radiative transfer modelling, once more adding an uncertainty to the observed abundances (Sect.\ \ref{sect:temp_dependence}).
	\\ \indent In order to better quantify uncertainties in resulting abundances, we perform a sensitivity analysis of the molecular envelope sizes of parent species and peak abundances of daughter species. Because of the degeneracy and multidimensionality of our parameter space, we separately consider the effect of an uncertainty in mass-loss rate (Sect.\ \ref{sect:sens_mdot}) and in temperature profile (Sect.\ \ref{sect:sens_eps}).}
	
	\begin{figure*}
		\centering
		\includegraphics[width=0.95\textwidth]{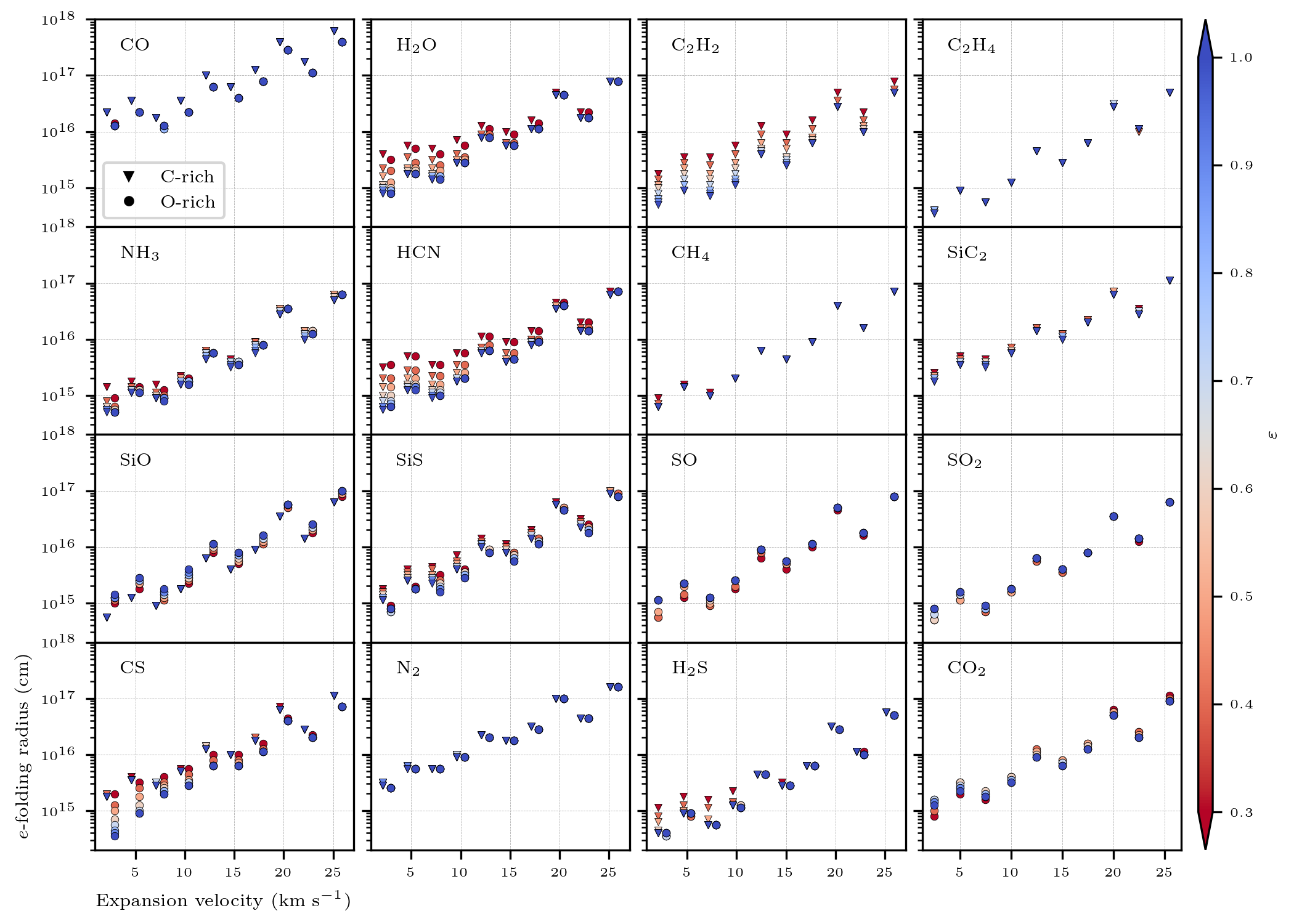}
		\caption{{Variation of the $e$-folding radii of parent species per expansion velocity for different C-rich (triangles) and O-rich (circles) models, with a density setup given by the green symbols in Fig.\ \ref{fig:grid}. The temperature profile of each model is set by $\Tstar = 2500\,$K and $\eps$ given by the colour (Eq.\ \ref{eq:temp_profile}). For better visibility and differentiation between C-rich and O-rich results, the symbols are centred around the corresponding expansion velocity for certain parents.}}
		\label{fig:sens_parents_eps}
	\end{figure*}
	\begin{figure*}
		\centering
		\includegraphics[width=0.7\textwidth]{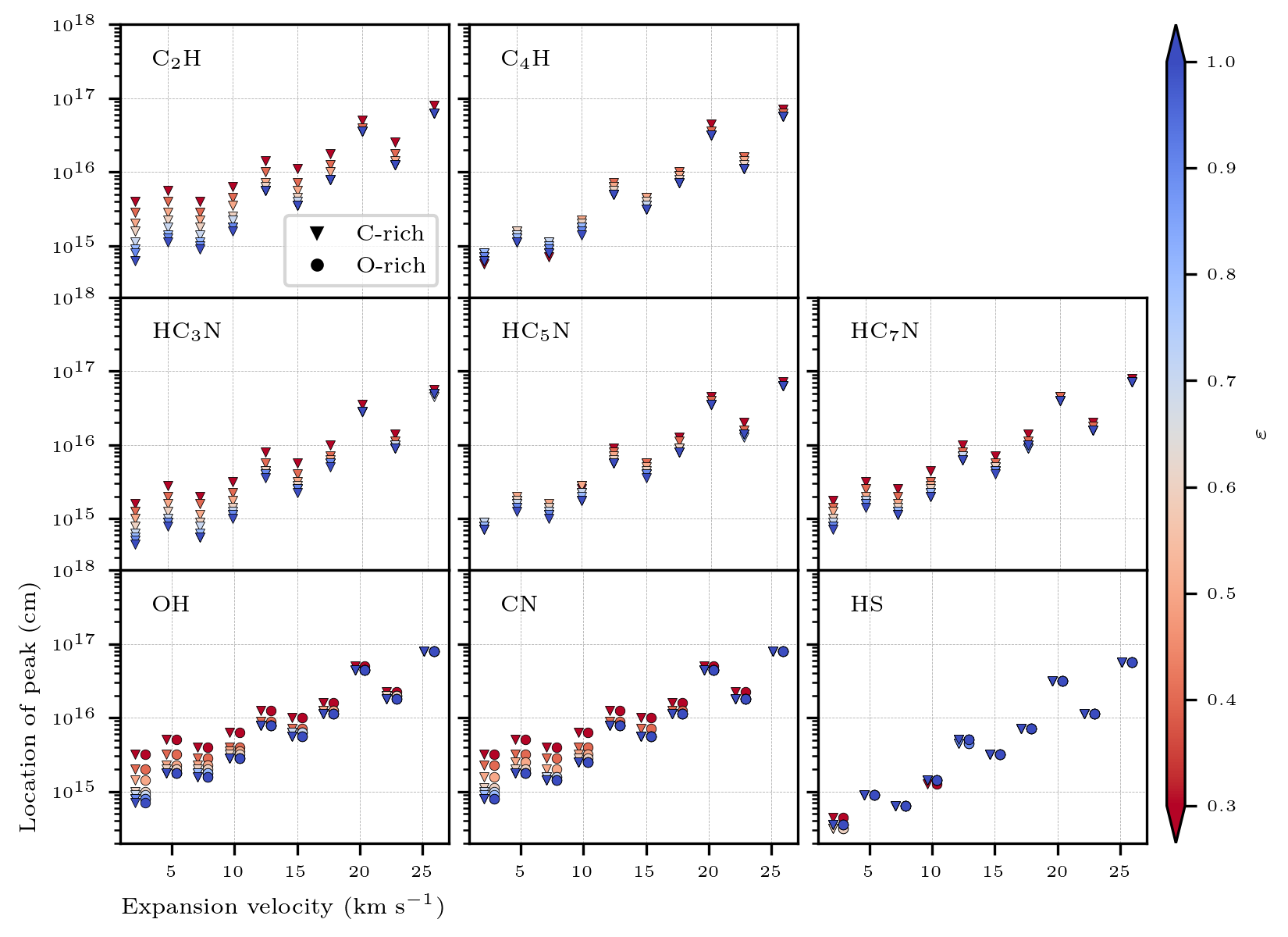}
		\caption{{Location of the peak in abundance profile of daughter species per expansion velocity for different C-rich (triangles) and O-rich (circles) models, with a density setup given by the green symbols in Fig.\ \ref{fig:grid}. The temperature profile of each model is set by $\Tstar = 2500\,$K and $\eps$ given by the colour (Eq.\ \ref{eq:temp_profile}). For better visibility and differentiation between C-rich and O-rich results, the symbols are centred around the corresponding expansion velocity for certain daughters. }}
		\label{fig:sens_daughters_r_eps}
	\end{figure*}
	\begin{figure*}
		\centering
		\includegraphics[width=0.7\textwidth]{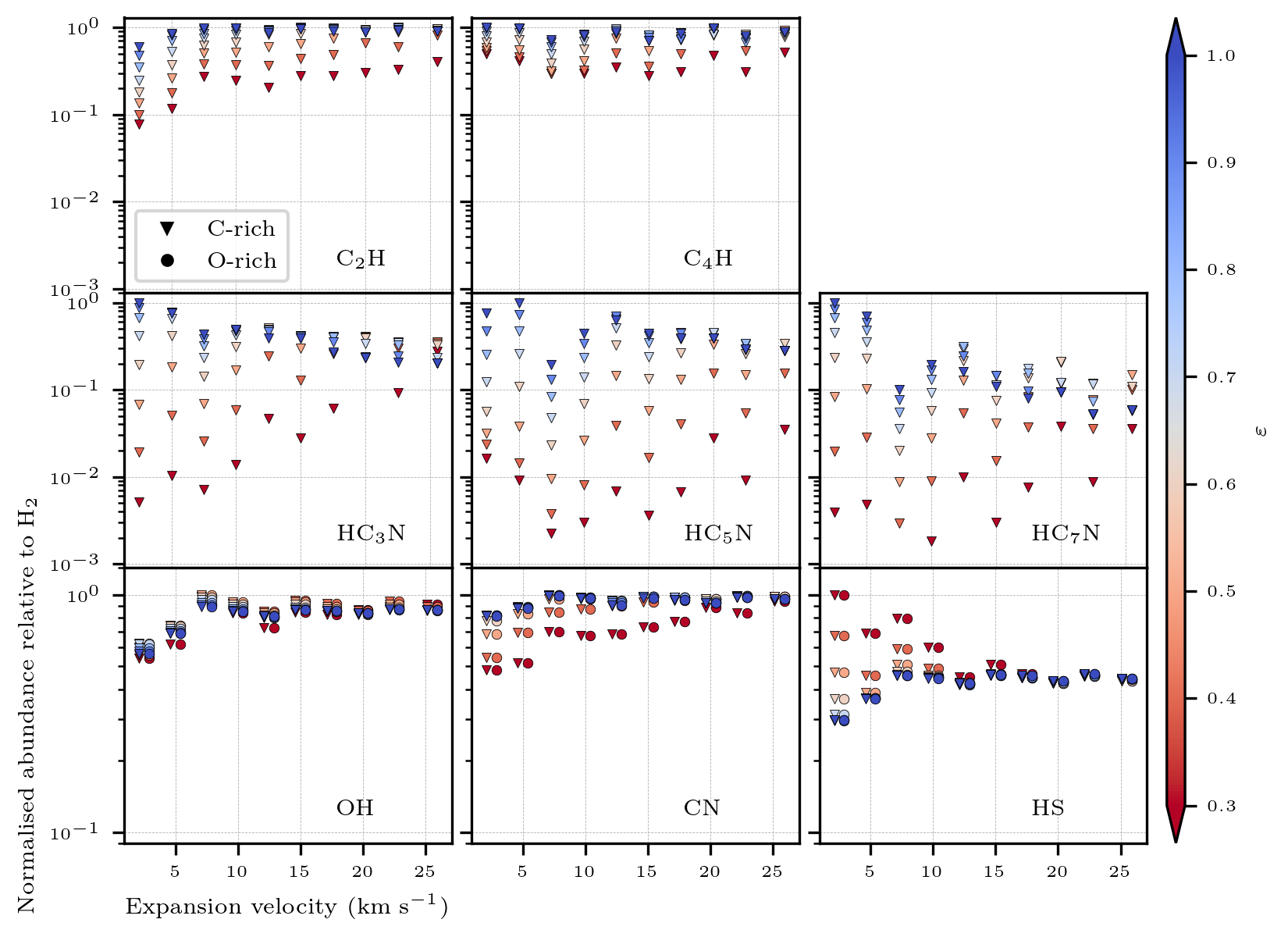}
		\caption{{Peak abundance relative to H$_2$, normalised to maximum abundance in the panel, of daughter species per expansions velocity for different C-rich (triangles) and O-rich (circles) models, with a density setup given by the green symbols in Fig.\ \ref{fig:grid}. The temperature profile of each model is set by $\Tstar = 2500\,$K and $\eps$ given by the colour (Eq.\ \ref{eq:temp_profile}). For better visibility and differentiation between C-rich and O-rich results, the symbols are centred around the corresponding expansion velocity for certain daughters. Maximum abundances relative to H$_2$: C$_2$H: $8.700\times10^{-6}$, C$_4$H: $6.188\times10^{-7}$, HC$_3$N: $2.727\times10^{-6}$, HC$_5$N: $3.456\times10^{-7}$, HC$_7$N: $2.492\times10^{-7}$, OH: $4.458\times10^{-5}$, CN: $6.879\times10^{-8}$, HS: $4.861\times10^{-6}$.}}
		\label{fig:sens_daughters_fracs_eps}
	\end{figure*}
	\subsection{Effect of uncertainty on mass-loss rate}\label{sect:sens_mdot}
	{In this Section, we consider models for different mass-loss rates per expansion velocity of our grid (Fig.\ \ref{fig:grid}) for a fixed, average temperature profile of $\Tstar = 2500\,$K and $\eps = 0.6$ \citep{Millar2004}. In Figs.\ \ref{fig:sens_parents}, \ref{fig:sens_daughters_r}, and \ref{fig:sens_daughters_fracs}, we demonstrate the effect of different mass-loss rates on the abundances of specific species. 
	\\ \indent The $e$-folding radii of the parent species of both C-rich and O-rich outflows are shown in Fig.\ \ref{fig:sens_parents} per expansion velocity, the mass-loss rate indicated by the colour. The steps in mass-loss rate in our grid approximately correspond to factors of 2. For an observationally determined expansion velocity and mass-loss rate of an AGB outflow, the uncertainty in envelope size of a given molecule can be estimated by considering subsequent vertical points above and below the model corresponding best to the observation, after appointing an uncertainty on the observationally determined mass-loss rate (since this is most often unknown).
	\\ \indent From Fig.\ \ref{fig:sens_parents}, we find that for a fixed uncertainty in mass-loss rate (i.e., a specific number of subsequent vertical points in the plot), the uncertainty range on the $e$-folding radius generally becomes larger for higher mass-loss rates. Fig.\ \ref{fig:sens_parents} shows that, over all, one may expect an uncertainty of about half an order of magnitude on the envelope size, if the uncertainty in mass-loss rate would be one order of magnitude.}
	\\ \indent {For commonly observed daughter species, the location of the peak abundance is shown in Fig.\ \ref{fig:sens_daughters_r} in a way analogous to the parents' envelope sizes. Fig.\ \ref{fig:sens_daughters_fracs} shows the peak abundance itself relative to H$_2$, normalised to the maximum abundance, given in the caption. The location of the peak of different daughters in the outflow shows a similar trend as the envelope sizes of the parents: at higher mass-loss rates, the uncertainty on the radius is generally larger, given a specific uncertainty on the mass-loss rate. Moreover, for some species, e.g. OH, CN, and HS, the abundance at the peak is more or less constant over varying mass-loss rates (see Fig.\ \ref{fig:sens_daughters_fracs}), while for others, e.g. the cyanopolyynes, the abundance at the peak changes by more than an order of magnitude. 
	}

	\subsection{Effect of uncertainty on temperature profile}\label{sect:sens_eps}
	{In this Section, we estimate the uncertainty ranges on the abundances due to an uncertainty in temperature profile. Since the exponent of the power law, $\eps$, has the most influence (Fig.\ \ref{fig:grid}, right panel), we only vary this parameter and fix the stellar temperature again at 2500\,K. To avoid crowding the figure, we used a subsample of the grid. Figs.\ \ref{fig:sens_parents_eps}, \ref{fig:sens_daughters_r_eps}, and \ref{fig:sens_daughters_fracs_eps} show the results per expansions velocity for different mass-loss rates, indicated in green in Fig.\ \ref{fig:grid}. 
	\\ \indent Fig.\ \ref{fig:sens_parents_eps} shows the variation in molecular envelope size of parent species due to different exponents $\eps$, indicated with the colour. The zig-zag trend is due to the difference in CO self-shielding at different outflow densities (Sect.\ \ref{sect:COselfshielding}), here ordered by the expansion velocity only. It is clear that certain parent species, e.g. CO, N$_2$, and C$_2$H$_4$, CH$_4$ and SiC$_2$ in the C-rich case, are more robust to changes in the temperature profile than others, e.g., H$_2$O, HCN, C$_2$H$_2$ in C-rich outflows, and CS in O-rich outflows. The latter are parents that are reformed directly or indirectly by reactions including energy barriers. Generally, an uncertainty on the mass-loss rate introduces a larger range on the envelope size than an uncertainty on the temperature exponent.
	\\ \indent The locations of the peak abundance and the peak abundance itself for certain daughter species are shown in Fig.\ \ref{fig:sens_daughters_r_eps} and \ref{fig:sens_daughters_fracs_eps}, respectively. Again, the range on these quantities is generally larger due to an uncertainty on the mass-loss rate than due to the uncertainty on the temperature profile.}

	\section{Comparison to observations} \label{sect:observations}
	{In this section, the findings of the models are compared to observed abundances in the outflow of AGB sources. However, a one-to-one comparison with abundances from specific sources is beyond the scope of this project. The majority of chemical species regularly observed in AGB outflows are species we consider to be parents (see e.g.\ \citealp{Gonzalez2003, Schoier2013, Decin2018}). Hence, this is an input of the models and, as such, variations in the abundance of these species in observed sources, cannot be analysed. Additionally, systematic observational studies of daughter species in multiple sources are presently rare. 
	\\ \indent This section is split between daughter and parent species. In Sect.\ \ref{sect:obs_daughters}, we compare our results for CN to the study of \cite{Bachiller1997}, and in Sect.\ \ref{sect:obs_parents_envs}, we elaborate on our relations of the envelope sizes and compare to similar relations extracted from observational studies.}
		
	\subsection{Daughter species}\label{sect:obs_daughters}
	{Valuable information about the physical parameters can be hidden in the abundance profiles of daughter species (Sect.\ \ref{sect:discussion}). Observing the appropriate set of daughter species, in combination with parents, can help the determination of the physical parameters of the outflow. In Appendix \ref{sect:detectability} we establish this as a proof-of-concept.
	\\ \indent \cite{Bachiller1997} performed a survey of CN in CSEs of 33 AGB sources. For the 26 C-rich sources, they found an average peak abundance of $1.9\times10^{-5}$, and for 7 O-rich source, an average abundance of $6.6\times10^{-5}$, both with respect to H$_2$. These abundances were compared to theoretical models of, e.g., \cite{MillarHerbst1994} and \cite{NejadMillar1988}. With an updated and larger chemical network compared to the previous studies, the average peak abundance of CN in our models is closer to the observed values. A comparison between the results can be found in Table \ref{tab:CN_obs}. Hence, this indicates that improving chemical models is necessary and helps to explain observed abundances.}
	\begin{table}
		\begin{center}
			\caption{Comparison of peak fractional abundances of CN with respect to H$_2$.}
			\begin{tabular}{   l  c  r  }
				\hline \hline \\[-2ex]
				 & Observed$^{a}$ & Modelled \\ \hline
				 C-rich & $1.9\times10^{-5}$& $\sim 8\times 10^{-6}$ $^{b}$ \\
				  & & $(1.09 \pm 0.03)\times 10^{-5}$ $^{d}$ \\
				 O-rich &$6.6\times10^{-8}$ & $4-40\times10^{-8}$  $^{c}$\\
				  & & $(6.1\pm0.3)\times10^{-8}	$ $^{d}$
				\\  \hline
			\end{tabular}
			\label{tab:CN_obs}
		\end{center}
	{\footnotesize {\textbf{Note.} $^a$ \protected \cite{Bachiller1997}, $^b$ \protected\cite{MillarHerbst1994}, $^c$ \protected\cite{NejadMillar1988}, $^d$ this work.} 
	}
	\end{table}
	\subsection{Parent's envelope sizes}\label{sect:obs_parents_envs}
	The $e$-folding radii of different species have been studied from an observational perspective, resulting in relations between outflow density and the $e$-folding radius for certain species. We compare the linear fits from the envelope sizes (Eq.\ \ref{eq:loglin_fit}, table \ref{tab:fit_env_ext_all}) of the different models to the relations found in the literature. The density parameters of the sources in the literature are shown in {Fig.\ \ref{fig:literature_envs}} and compared with our parameter space.
	\\ \indent \cite{NetzerKnapp1987} investigated H$_2$O and OH in O-rich circumstellar envelopes by means of their maser emission on theoretical grounds, using a grid of models with mass-loss rates ranging from $10^{-7}$ to $10^{-4}\,\Msolyr$ and expansions velocities between 5 and 40\,$\kms$. \cite{Maercker2008} refined the application of this to observational data, and came to a relation between the $e$-folding radius of H$_2$O and outflow density, as a function of $\Mdot/\vexp$:
	\begin{equation}\label{eq:efolding_H2O}
		R_e({\rm H_2O}) =5.4\times 10^{16}\left(\frac{\Mdot}{10^{-5}}\right)^{0.7}\left(\vexp\right)^{-0.4}
	\end{equation}
	with $\Mdot$ in $\Msolyr$, $\vexp$ in $\kms$, and $R_e$ in cm. The results for H$_2$O are shown in Fig.\ \ref{fig:H2O_env}, where we fitted Eq.\ (\ref{eq:efolding_H2O}) with the linear relation from Eq.\ (\ref{eq:loglin_fit}), producing the dashed line. The envelope sizes found in the present study for both chemistry types are systematically larger than the relation derived by \cite{Maercker2008}. We note that \cite{Maercker2016} could not find a strong constraint for $R_e$ from the Herschel HIFI observational data, since the detected H$_2$O lines do not probe the outer envelope well and there was a degeneracy between abundance and envelope size. Hence, only additional data from H$_2$O lines would enable us to observe accurately its envelope, especially with the help of interferometry to spatially resolve low energy lines.
		\begin{figure}
		\centering
		\includegraphics[width=0.49\textwidth]{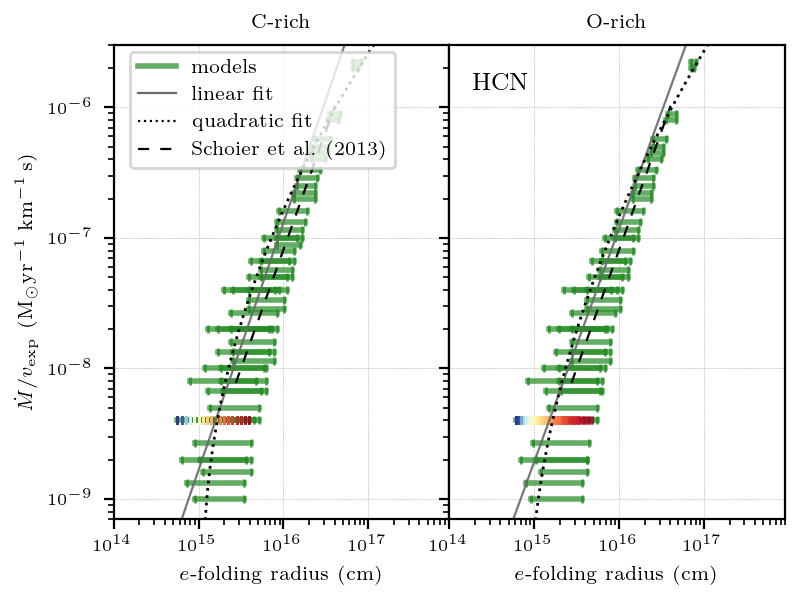}
		\caption{$e$-folding radii of HCN for the C-rich (\textit{left}) and O-rich (\textit{right}) models in green. For one density the reference temperature (see Sect.\ \ref{sect:results}) is indicated with the colour bar; the blue side indicates the cold outflows ($\eps=1.0$) and the red side the warm outflows ($\eps = 0.3$). The full grey line represents the linear fit to the data (Eq.\ \ref{eq:loglin_fit}, Table \ref{tab:fit_env_ext_all}), the dotted line represents the linear fit (Eq.\ \ref{eq:logq_fit}, Table \ref{tab:fit_quadr_env_ext_all}). The black dashed line follows Eq.\ (\ref{eq:efolding_HCN}) {for the density range of the source used by \protect\cite{Schoier2013}}.}
		\label{fig:HCN_env}
	\end{figure}
	\begin{figure}
		\centering
		\includegraphics[width=0.49\textwidth]{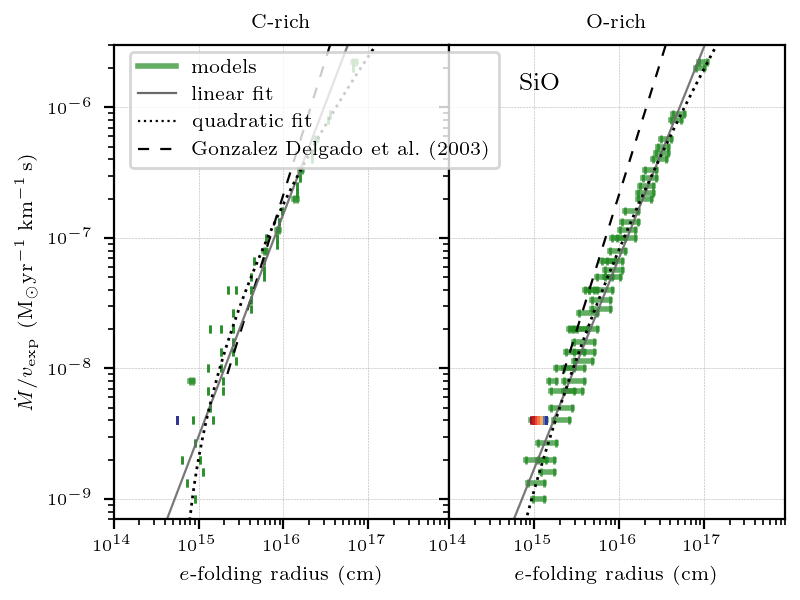}
		\caption{$e$-folding radii of SiO for the C-rich (\textit{left}) and O-rich (\textit{right}) models in green. For one density the reference temperature (see Sect.\ \ref{sect:results}) is indicated with the colour bar; the blue side indicates the cold outflows ($\eps=1.0$) and the red side the warm outflows ($\eps = 0.3$). The full grey line represents the linear fit to the data (Eq.\ \ref{eq:loglin_fit}, Table \ref{tab:fit_env_ext_all}), the dotted line represents the linear fit (Eq.\ \ref{eq:logq_fit}, Table \ref{tab:fit_quadr_env_ext_all}). The black dashed line follows Eq.\ (\ref{eq:efolding_SiO}) {for the density range of the source used by \protect\cite{Gonzalez2003}}.}
		\label{fig:SiO_env}
	\end{figure} 
	\\ \indent HCN is to interest in O-rich outflows, as it highlights the non-equilibrium character of the chemistry in AGB outflows. \cite{Schoier2013} used radiative transfer models to determine envelope sizes from the observed thermal line emission of the HCN envelope, assuming Gaussian abundance profiles. They found the following relation from a sample of about 20 sources including C-rich, O-rich, and S-type stars:
	\begin{equation}\label{eq:efolding_HCN}
	\log R_e({\rm HCN}) = (19.9\pm 0.6) + (0.55\pm 0.09)\,\log\left(\frac{\Mdot}{\vexp}\right).
	\end{equation}
	Fig.\ \ref{fig:HCN_env} shows the results for the $e$-folding radius of HCN. The fit to the envelope sizes of our models corresponds well with the relation found by \cite{Schoier2013} (Eq.\ \ref{eq:efolding_HCN}), especially for higher outflow densities. This is the case for both the C-rich and O-rich models. This implies that the chemical network used contains the most relevant reactions involving HCN.
	\\ \indent \cite{Gonzalez2003} investigated the extent of the SiO envelope observationally, since this species is of interest for dust formation in O-rich CSEs. By fitting radiative transfer models to observed SiO lines in a large sample ($\sim 70$) of O-rich outflows, they found a lower limit for the size of the SiO envelope to be
	\begin{equation}\label{eq:efolding_SiO}
		\log R_e({\rm SiO}) = 19.2+0.48\log\left(\frac{\Mdot}{\vexp}\right), 
	\end{equation}
	as a function of $\Mdot/\vexp$. \cite{Schoier2006} found this relation to be also valid for a small sample of C-rich AGB stars, within the observational uncertainties. \cite{Ramstedt2009} confirm, with a similar study of S-type stars, that the envelope size of SiO is not very sensitive to the C/O-ratio in the outflow. The results for SiO are given in Fig.\ \ref{fig:SiO_env}. Our modelled envelope sizes agree well with observed envelope sizes of SiO (Eq. \ref{eq:efolding_SiO}), for both the O-rich and C-rich models, validating our models with the observational studies. However, in the O-rich, we see that for higher densities, the models start to diverge from Eq. (\ref{eq:efolding_SiO}). SiO is often not detected up to $10^{17}\,{\rm cm}$ away from the star in observations, possibly due to depletion onto dust \citep{VandeSande2019, Massalkhi2019, Massalkhi2020}. \\
	\begin{figure}
		\centering
		\includegraphics[width=0.49\textwidth]{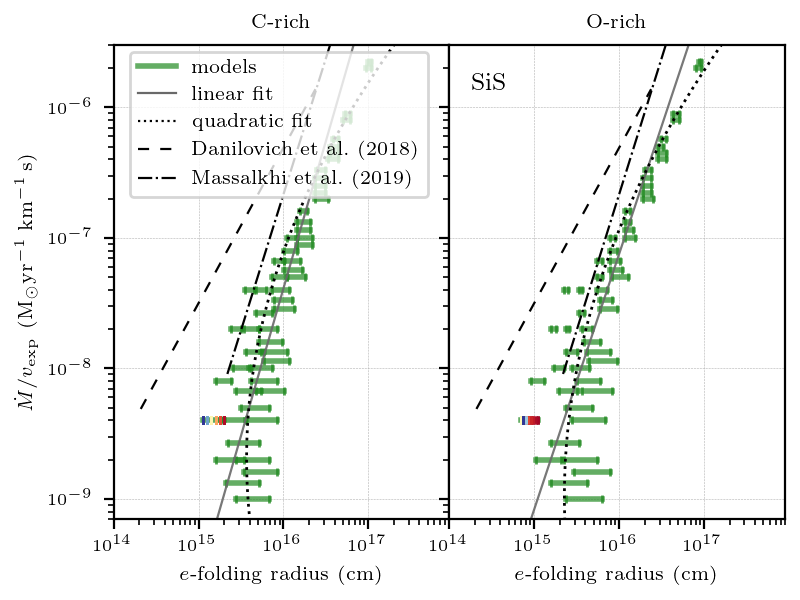}
		\caption{$e$-folding radii of SiS for the C-rich (\textit{left}) and O-rich (\textit{right}) models in green. For one density the reference temperature (see Sect.\ \ref{sect:results}) is indicated with the colour bar; the blue side indicates the cold outflows ($\eps=1.0$) and the red side the warm outflows ($\eps = 0.3$).The full grey line represents the linear fit to the data (Eq.\ \ref{eq:loglin_fit}, Table \ref{tab:fit_env_ext_all}), the dotted line represents the linear fit (Eq.\ \ref{eq:logq_fit}, Table \ref{tab:fit_quadr_env_ext_all}). The black dashed line follows Eq.\ (\ref{eq:efolding_SiS}) the dashed-dotted black line Eq.\ (\ref{eq:efolding_SiO}), {for the density ranges of the source used by \protect\cite{Danilovich2018} and \protect\cite{Massalkhi2019}, respectively.}}
		\label{fig:SiS_env}
	\end{figure}

	\indent The photodissociation rate of the parent species SiS has not been determined experimentally, and generally the assumption is made that it behaves similarly to SiO with respect to photodissociation \citep{VanDishoeck1988,Wirsich1994}. Hence,  \cite{Massalkhi2019} used Eq.\ (\ref{eq:efolding_SiO}) for the envelope sizes of SiS in their study. In our chemical network, this assumption is also made: the UMIST database uses an old SiO rate for the photodissociation of SiS \citep{McElroy2013}. In Fig.\ \ref{fig:SiS_env} we see that the envelope sizes of SiS agree decently with the relation assumed by \cite{Massalkhi2019} (dashed-dotted line), albeit the modelled envelopes are systematically slightly larger. However, \cite{Danilovich2018} found a different description for the envelope size of SiS, when they analysed a smaller sample of AGB outflows (containing C-rich, but mostly O-rich outflows), and using more lines compared to \cite{Massalkhi2019}. Their data fitted the relation
	\begin{equation}\label{eq:efolding_SiS}
		\log R_e({\rm SiS}) = (21.3\pm 0.2) +(0.84\pm 0.03)\,\log\left(\frac{\Mdot}{\vexp}\right), 
	\end{equation}
	adopting a similar method as \cite{Schoier2013}. 
	This is given by the dashed line in Fig.\ \ref{fig:SiS_env}. Since our envelope sizes of SiS differ considerably from the relation found by \cite{Danilovich2018}, we conclude that the assumption of the photodissociation of SiS behaving similar to SiO is incorrect, and thus that important chemical data is missing. 
		\begin{figure}
		\centering
		\includegraphics[width=0.49\textwidth]{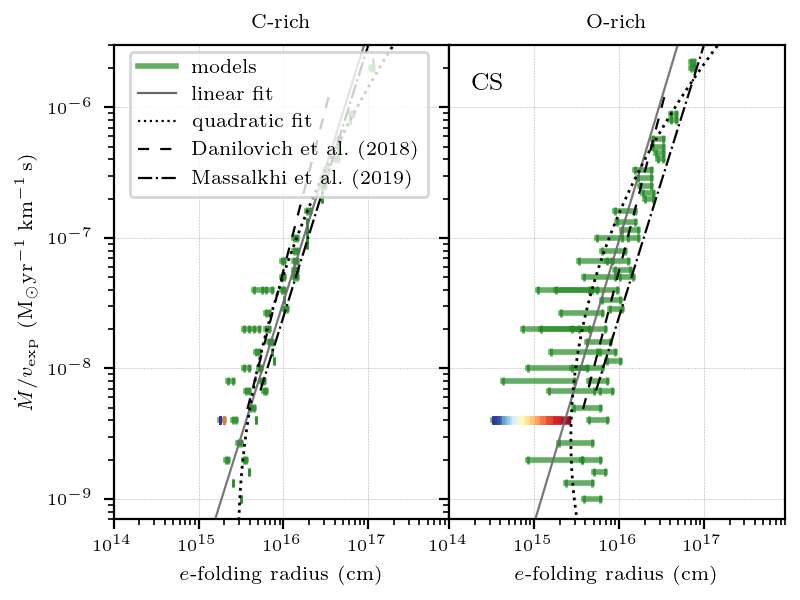}
		\caption{$e$-folding radii of CS for the C-rich (\textit{left}) and O-rich (\textit{right}) models in green. For one density the reference temperature (see Sect.\ \ref{sect:results}) is indicated with the colour bar; the blue side indicates the cold outflows ($\eps=1.0$) and the red side the warm outflows ($\eps = 0.3$). The full grey line represents the linear fit to the data (Eq.\ \ref{eq:loglin_fit}, Table \ref{tab:fit_env_ext_all}), the dotted line represents the linear fit (Eq.\ \ref{eq:logq_fit}, Table \ref{tab:fit_quadr_env_ext_all}). The black dashed line follows Eq.\ (\ref{eq:efolding_CS_Danilovich2018}), the dashed-dotted black line Eq.\ (\ref{eq:efolding_CS_Massalkhi2019}), {for the density ranges of the source used by \protect\cite{Danilovich2018} and \protect\cite{Massalkhi2019}, respectively.}}
		\label{fig:CS_env}
	\end{figure}
	\\ \indent The parent CS is also examined in the same study of \cite{Danilovich2018}. They found that the $e$-folding radius of CS follows
	\begin{equation}\label{eq:efolding_CS_Danilovich2018}
		\log R_e({\rm CS}) = (18.9\pm 0.2) +(0.40\pm 0.03)\,\log\left(\frac{\Mdot}{\vexp}\right),
	\end{equation} 
	as a function of $\Mdot/\vexp$. \cite{Massalkhi2019} applied a different relation for their C-rich sample, namely
	\begin{equation}\label{eq:efolding_CS_Massalkhi2019}
		\log R_e({\rm CS}) = 19.65+ 0.48\log\left(\frac{\Mdot}{\vexp}\right).
	\end{equation}
	They started from the relation found for SiO by \cite{Gonzalez2003} (Eq.\ \ref{eq:efolding_SiO}), but adopting a larger radial extent due to anomalously high CS abundances for some sources in their sample.
	Fig.\ \ref{fig:CS_env} shows the results for CS. Although both studies were done with a different set of stars, the two relations differ only at higher densities. For the C-rich chemistry, our models follow Eqs.\ (\ref{eq:efolding_CS_Danilovich2018}) and (\ref{eq:efolding_CS_Massalkhi2019}) well, but the correspondence with \cite{Massalkhi2019} (Eq.\ \ref{eq:efolding_CS_Massalkhi2019}) is better, which is expected since their sample contains C-rich stars only, contrary to \cite{Danilovich2018} (Eq.\ \ref{eq:efolding_CS_Danilovich2018}). On the other hand, for the O-rich chemistry, the envelope sizes of CS diverge from both relations, especially at lower density, indicating there could be chemistry missing in our network, possibly related to shocks in the medium.

	\section{Conclusions}\label{sect:conclusion}
	We have performed the first sensitivity study of the chemistry taking place in circumstellar envelopes of AGB stars, with respect to the physical environment of these CSEs. To this end, we calculated 1D chemical kinetics models of smooth outflows including gas-phase chemistry only, and analysed the resulting abundance profiles of parent and daughter species as a function of temperature and density. 
	\\ \indent From the chemistry point-of-view, we find that the abundance profiles depend on the density of the outflow due to the extinction-dependence of the photodissociation rate. Therefore, when the density is high, parent species are only photodissociated further out in the wind, shifting the abundance peaks of daughter species accordingly further away from the star. {Abundance profiles depend on the specific temperature profile in the outflow when the main chemical reaction pathway involves an energy barrier.} The specific temperature dependence of a species can be inherited by subsequent generations of daughter species. We determined the dependence of the outflow density of the envelope size of parent species, $R_e$, by fitting the logarithm of $R_e$ as a function of the logarithm of the density to a linear relation {and to a quadratic relation. It is found that the linear relation is accurate to describe the envelope sizes up to first order.} Further, we analysed the envelope size of CO in detail, and compared with the studies by \cite{Groenewegen2017} and \cite{Saberi2019} with respect to the CO self-shielding. The single-band approximation used in our models is found to be sufficiently accurate for this study of the chemistry in CSEs. {We compared linear relations of the modelled molecular envelope sizes} to similar relations found in the literature, which are mainly derived from observational studies. For most of the parent species our models agreed well with the literature relations. We found a significant difference between our model results and the literature relation for SiS, further emphasising the need for an accurate determination of its photodissociation rate.
	\\ \indent {The results presented here can aid observational studies to determine uncertainties on molecular abundances and envelope sizes, given a certain uncertainty on the physical parameters of the outflow. The uncertainties on the envelope sizes of parent species and locations of peak abundances of daughter species is generally found to be about half an order of magnitude, if the uncertainty on the mass-loss rate would be an order of magnitude. Depending on the density, this uncertainty increases by several factors, due to the uncertainty on the temperature profile. }

	\section*{Acknowledgements}
	The authors are grateful to the referee, who helped improving our manuscript. S.M.\ and L.D.\ acknowledge support from the ERC consolidator grant 646758 AEROSOL and from the Research Foundation Flanders (FWO) grant G099720N. M.V.d.S.\ is supported by the European Union's Horizon 2020 research and innovation programme under the Marie Sk\l odowska-Curie grant agreement No 882991. T.D.\ acknowledges supported from the Research Foundation Flanders (FWO) through grant 12N9920N and is supported in part by the Australian Research Council through a Discovery Early Career Researcher Award (DE230100183). F.D.C.\ is supported by a junior postdoctoral fellowship of the Research Foundation Flanders (FWO) grant nr. 1253223N.

	\section*{Data Availability}
	The chemical reaction network, chemical kinetics model, and all data underlying this article are available on request to the corresponding author.

	
	
	\bibliographystyle{mnras}
	\bibliography{references} 
	
	

	\appendix
	\section{UV radiation field \& outflow opacity}\label{sect:photodiss}
	{The chemical kinetics models adopts the interstellar Draine UV field \citep{Draine1978} to calculate the photodissociation rate of species. When this radiation penetrates the outflow, the radiation is extinguished with an opacity
	\begin{equation}\label{eq:opacity}
		\kappa = 1.086\frac{2[A_{UV}/A_V]}{1.87\times 10^{21}}\frac{1}{\mu m_H} \approx 1200\,{\rm cm^2\, g^{-1}},
	\end{equation}
	where $\mu = 2.68$ is the mean molecular mass of the outflow relative to H$_2$, including He (see Table \ref{tab:parents}), $m_H$ is the atomic mass unit. The prefactor 1.086 is needed for the conversion from extinction to optical depth. We assume that the extinction is equal to that of the ISM of $1.87\times 10^{21}$ atoms cm$^{-2}$ mag$^{-1}$ \citep{Cardelli1989}, and is scaled to the ratio of UV and visual extinction $A_{UV}/A_V=4.65$ \citep{Nejad1984}. 	}
	\section{Effect on potential detectability of species}\label{sect:detectability}
	In this Section, {we demonstrate that abundance profiles of specific daughter species can help constrain the temperature profile of the AGB outflow. We consider the column density as a rough indicator of potential detectability. Note that this is a proof-of-concept, since column density does not relate linearly to observability.} Further, we discuss the change in potential detectability of specific species caused by a different temperature profile or outflow density. 
	\subsection{Specific examples of daughters} \label{sect:detect_examples}
	\begin{figure}
		\centering
		\includegraphics[width=0.49\textwidth]{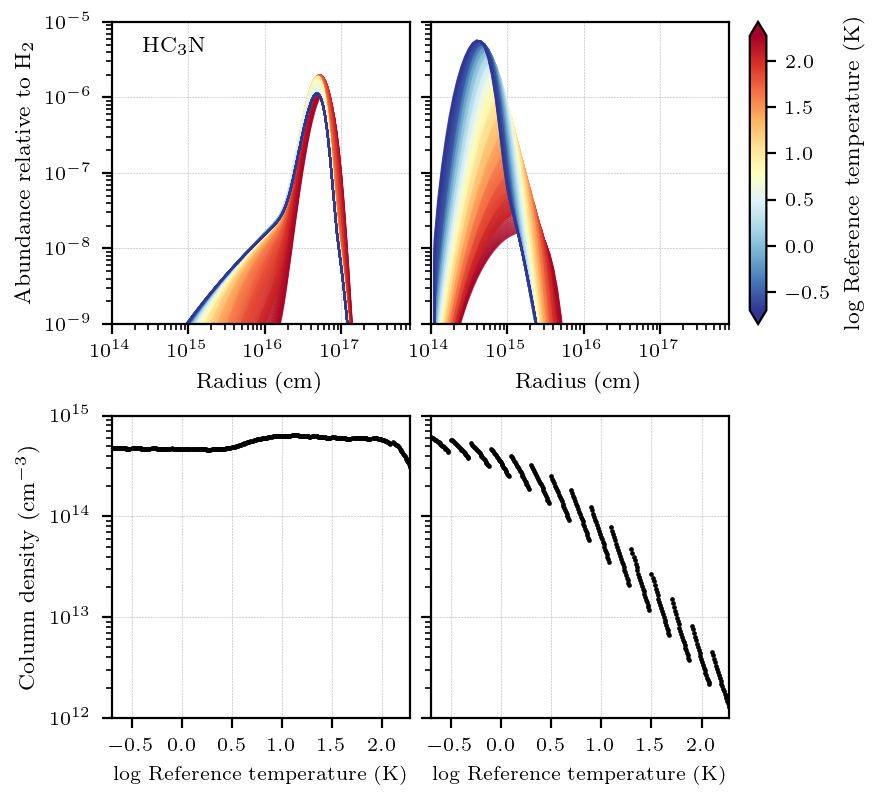}
		\caption{Fractional abundance profile and column density of {HC$_3$N} in a C-rich model, for a high and low outflow density: $(\vexp \ [{\rm \kms}],\Mdot \ [{\rm \Msolyr}])$: \textit{left} (25.0,\ $5\times10^{-5}$) and \textit{right} (2.5,\ $10^{-8}$).}
		\label{fig:C_HC3N}
	\end{figure}
\begin{figure}
		\centering
		\includegraphics[width=0.49\textwidth]{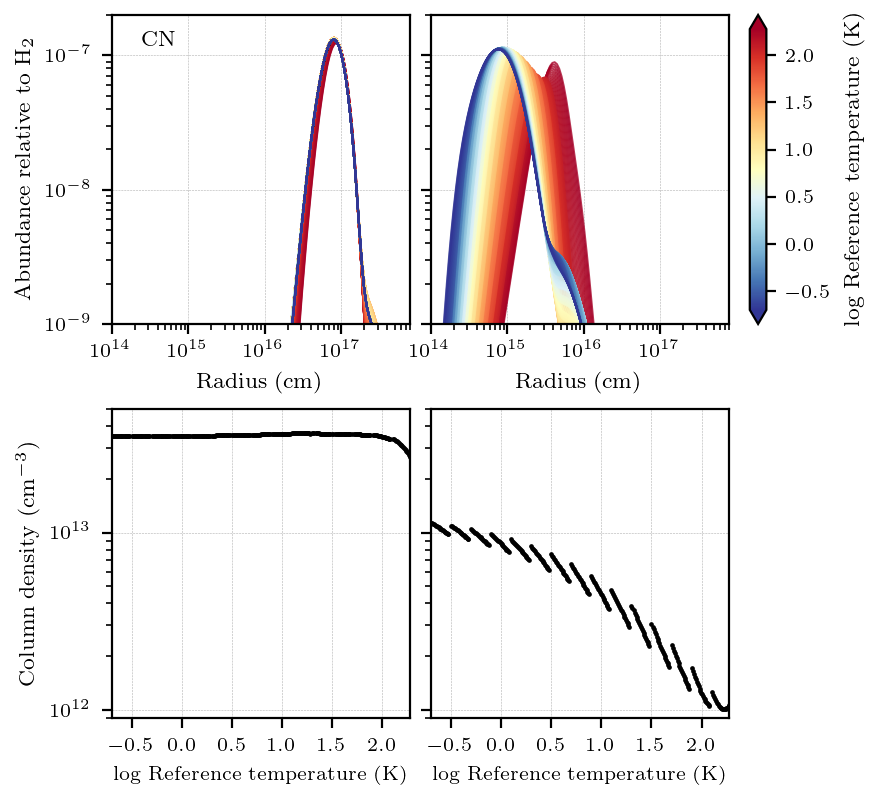}
		\caption{Fractional abundance profile and column density of {CN} in an O-rich model,for a high and low outflow density: $(\vexp \ [{\rm \kms}],\Mdot \ [{\rm \Msolyr}])$: \textit{left} (25.0,\ $5\times10^{-5}$) and \textit{right} (2.5,\ $10^{-8}$).}
		\label{fig:O_CN}
	\end{figure}
\begin{figure}
		\centering
		\includegraphics[width=0.49\textwidth]{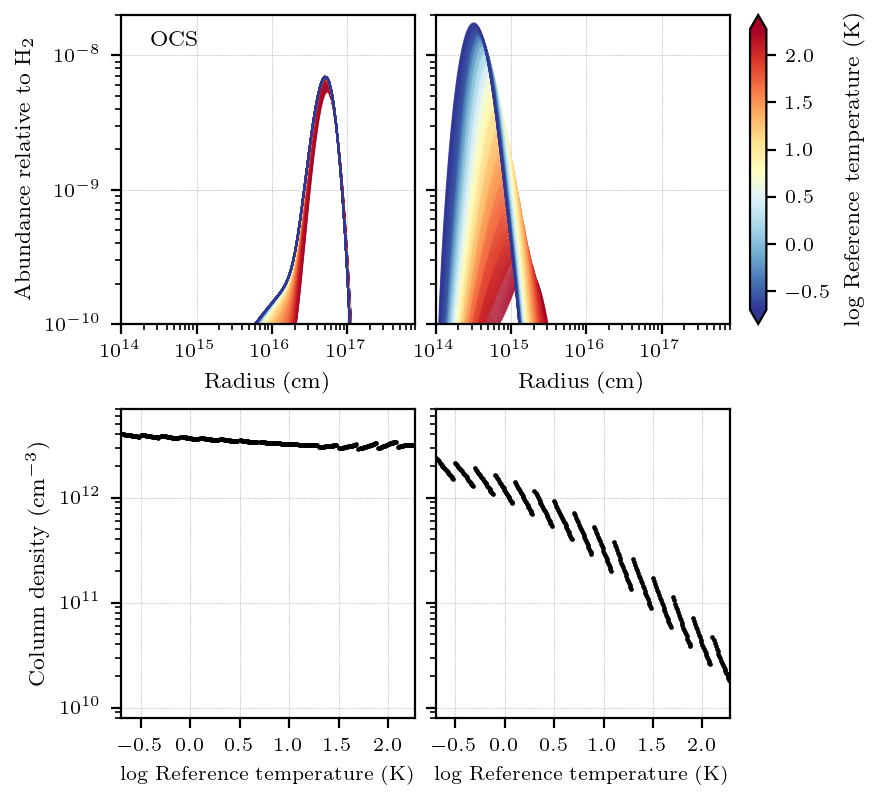}
		\caption{Fractional abundance profile and column density of OCS in an O-rich model,for a high and low outflow density: $(\vexp \ [{\rm \kms}],\Mdot \ [{\rm \Msolyr}])$: \textit{left} (25.0,\ $5\times10^{-5}$) and \textit{right} (2.5,\ $10^{-8}$).}
		\label{fig:O_OCS}
	\end{figure}

	In Fig.\ \ref{fig:C_HC3N}, the abundance profile of {HC$_3$N, the smallest cyanopolyyne, in a C-rich outflow is shown} together with its column density as a function of reference temperature, for a high and low outflow density. The effect of different $\Tstar$ and $\eps$ (Eq.\ \ref{eq:temp_profile}) values becomes visible in the trend of the column density: we distinguish distinct groups of models, which represent models of the same $\eps$ value, and different $\Tstar$ values within these groups. The behaviour of the column density with temperature is different for the two outflow densities. More specifically, for the high density the column density increases with temperature (bottom left panel), and it decreases with temperature for low outflow densities (bottom right panel). This is explained by the following: {HC$_3$N is formed through reaction (\ref{eq:cyanopolyynes}).
	At high density, the abundance profiles of CN and C$_2$H$_2$ are less sensitive to temperature compared to low density (see Figs.\ \ref{fig:C_C2H2_C2H} and \ref{fig:C_cyanopolyynes}, and Sect.\ \ref{sect:dens_dependence}). Therefore, the abundance of HC$_3$N does not depend on temperature and its column density stays roughly constant for a varying reference temperature. However, for a low outflow density, the abundance profiles of CN and C$_2$H$_2$ depend more strongly on temperature. In this case, HC$_3$N inherits the profile shapes of CN and C$_2$H$_2$, resulting in a larger abundance for cooler outflows. The abundances of CN and C$_2$H$_2$ are higher closer to the star for cooler outflows, this is also the case for HC$_3$N. Accordingly, the column density decreases with increasing reference temperature. Therefore, the abundance of HC$_3$N throughout the outflow can constrain the physical parameters, such as the temperature profile, of the outflow. A similar reasoning holds for further generations of cyanopolyynes.}
	\\ \indent {In O-rich outflows, the column density of daughter species such as OH and CN show a similar dependence on reference temperature. Fig.\ \ref{fig:O_CN} shows the abundance profile and column density of CN. The origin of their behaviour is analogous to HC$_3$N, as they are direct daughter species of HCN and H$_2$O, respectively.}
\begin{figure}
	\centering
	\includegraphics[width=0.49\textwidth]{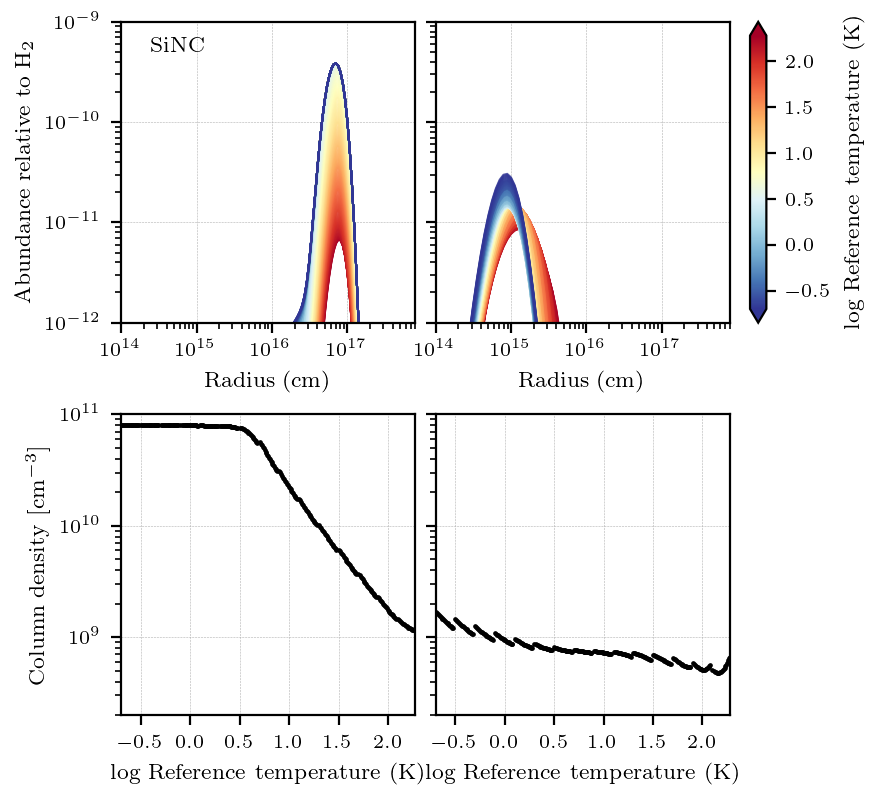}
	\caption{Fractional abundance profile and column density of SiNC in a C-rich model, for high and low outflow density: for a high and low outflow density: $(\vexp,\Mdot)$ in ${\rm(\kms,\Msolyr)}$: \textit{left} (25.0,\ $5\times10^{-5}$) and \textit{right} (2.5,\ $10^{-8}$). }
	\label{fig:C_SiNC}
\end{figure}
	\\ \indent Certain temperature profiles can lead to the column density exceeding a detectability threshold for specific daughter species. As an example, we take the case of SiNC in a C-rich outflow, e.g., observed in CW Leo \citep{Guelin2004}. 
	{At low density (Fig.\ \ref{fig:C_SiNC}, right panels) the abundance and column density of SiNC is low, around $10^9\,$cm$^{-3}$, and therefore, detecting SiNC is difficult or even impossible. However, SiNC can be observable in dense outflows, since in this case the abundance and column density are much higher. This is for example the case for the C-rich AGB star CW Leo \citep{Guelin2004}}. Moreover, for a higher outflow density the species is photodissociated further out in the outflow. Consequently, SiNC exists longer in the outflow, and this contributes to the column density. The highest abundance is found for models with a low reference temperature (high $\eps$, blue curves). 
	Hence, {when species that show a similar behaviour as SiNC, are observed in C-rich AGB outflows,} one may assume the density to be high and the temperature profile of the outflow to be rather steep ($\eps\gtrsim0.6$, Eq.\ \ref{eq:temp_profile}). 
	\\ \indent The species OCS has never been observed in AGB outflows to date, only in {other} post-main sequence objects (e.g., \citealp{Morris1987}), despite its importance to S-bearing chemistry (Sect.\ \ref{sect:temp_dependence}). Fig.\ \ref{fig:O_OCS} shows the abundance profiles of OCS and column density in an O-rich outflow. The column density is relatively high in both high and low density outflows, so based on this rough argument, we find that OCS is theoretically detectable. However, for a low density, the column density decreases with temperature, leading to a lower likelihood of its detection in low temperature outflows. 

	\subsection{Chemical thermometer}
	We have found that the temperature profile of the CSE generally has a crucial effect on the abundance profiles of the species in the outflow. Though, observationally speaking, constraining the temperature profile accurately is not straightforward. Generally, the kinetic temperature profile is retrieved from observations alongside synthetic line profiles with radiative transfer modelling, where an equation governs the energy balance between heating and cooling processes. Another way is to assume a power law for the temperature profile in the radiative transfer modelling. These methods make that a number of assumptions go into the modelling process. 
	\\ \indent The results from this study can therefore be used as an aid to observationally constrain temperature profiles of CSEs. It is best to use, as a so-called ``chemical thermometer'', a combination species. More specifically, we propose using parent-daughter pairs whose abundances largely depend on temperature, combined with species whose abundances are rather independent of temperature. For the former set of species, the location of the parents' destruction in the outflow, and the location of the daughters' peak, are dependent on the temperature profile (Sect.\ \ref{sect:temp_dependence}). The latter set of species is needed to rule out degeneracy in abundances due to a different outflow density. However, this technique will work better for sources that have a low outflow density, since the temperature dependency of the abundance profiles is more prominent in these type of outflows.
	\\ \indent  For example, in O-rich, low-density outflows we find that the following parent-daughter pairs make suitable chemical thermometers: HCN-CN and NH$_3$-NH$_2$. Also, the parent CS and daughter OH are suitable species. When the outflow density is higher, only OH and the HCN-CN pair remain somewhat temperature dependent. For all outflow densities, we propose to use CO and H$_2$S to constrain the density of the source. In C-rich outflows, when the outflow density is low, we find that the pairs C$_2$H$_2$-C$_2$H,  HCN-CN, and, H$_2$S-HS are suited as chemical thermometers. Also, the daughter species OH and HC$_3$N can be used to constrain the temperature profile. For higher outflow density sources, H$_2$S-HS quickly becomes rather independent of temperature, and hence not suited any more. To constrain the density of the outflow itself, we propose using species such as CO and CS.
	\\ \indent However, we note that these sets of suitable species heavily depend on the chosen parent species in the modelling process and on the physical parameters of the observed source. Therefore, it is difficult to pin down a general set of species suited for this purpose.
	
	\section{Additional figures}\label{sect:extrafigs}
	\begin{figure*}
		\centering
		\includegraphics[width=0.70\textwidth]{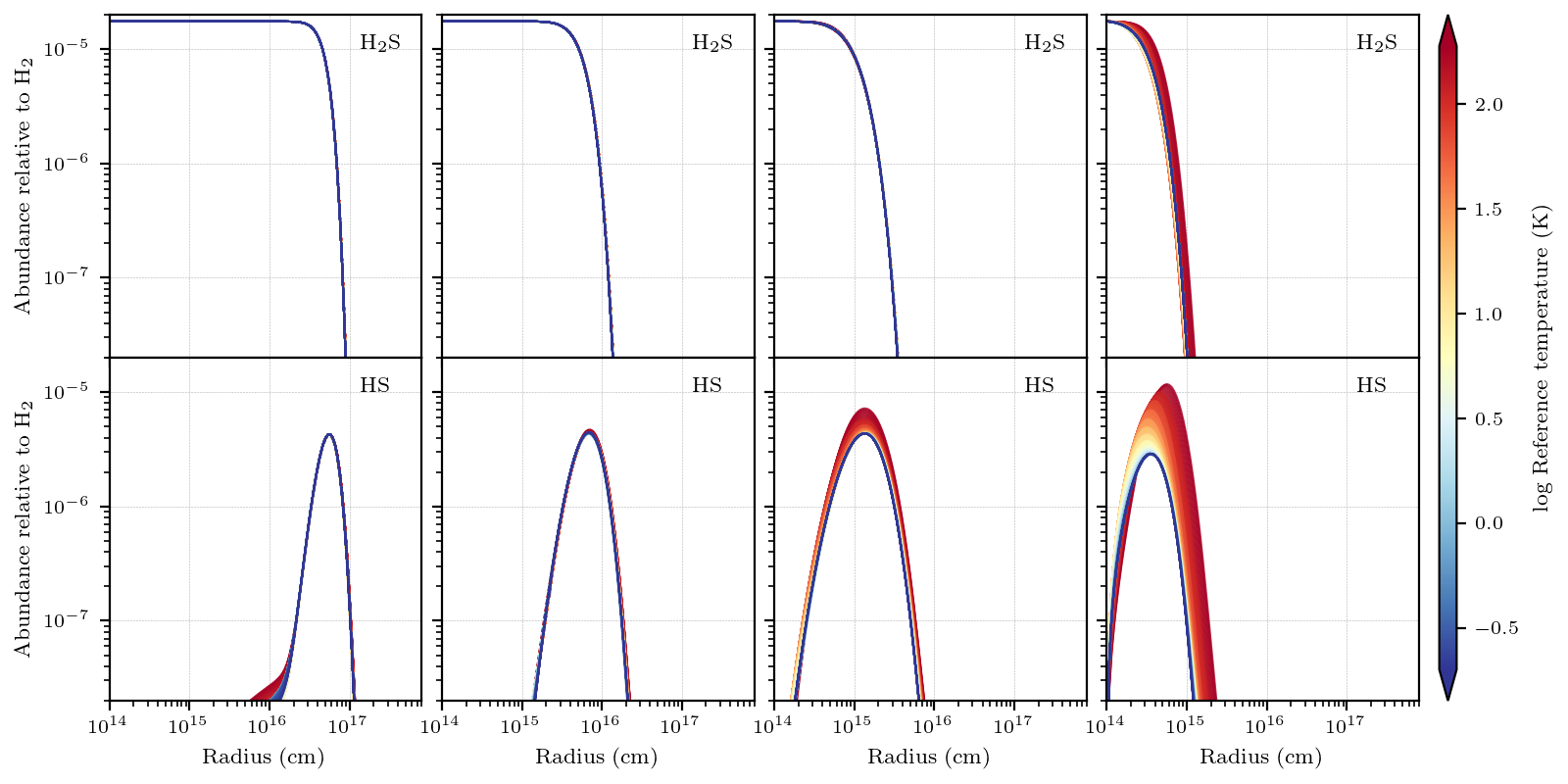}
		\caption{Fractional abundance profiles of H$_2$S and HS in O-rich outflows, going from high to low outflow density. From left to right $(\vexp \ [{\rm \kms}],\Mdot \ [{\rm \Msolyr}])$: (25.0,\ $5\times10^{-5}$), (17.5,\ $2\times10^{-6}$), (10.0,\ $10^{-7}$), (2.5,\ $10^{-8}$).}
		\label{fig:O_H2S_HS}
	\end{figure*}
	\begin{figure*}
		\centering
		\includegraphics[width=0.6\textwidth]{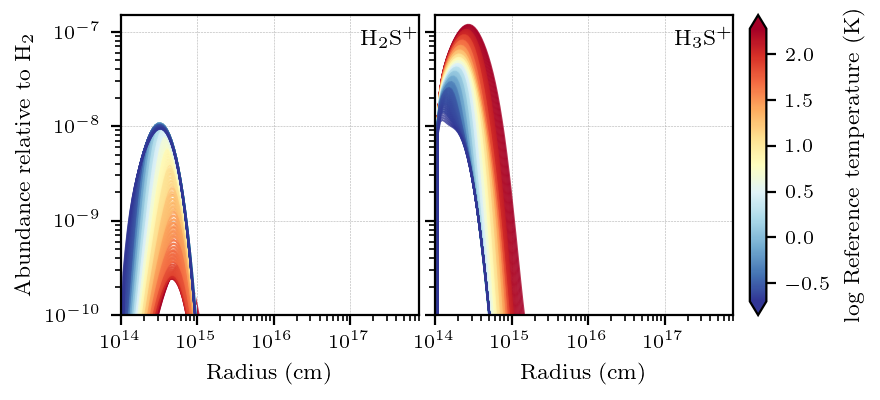}
		\caption{Fractional abundance profiles of H$_2$S$^+$ and H$_2$S$^+$ in an O-rich outflow, for high and low outflow density. From left to right $(\vexp \ [{\rm \kms}],\Mdot \ [{\rm \Msolyr}])$: (25.0,\ $5\times10^{-5}$) and (2.5,\ $10^{-8}$).}
		\label{fig:O_H2S+_H3S+}
	\end{figure*}
	\begin{figure*}
		\centering
		\includegraphics[width=0.7\textwidth]{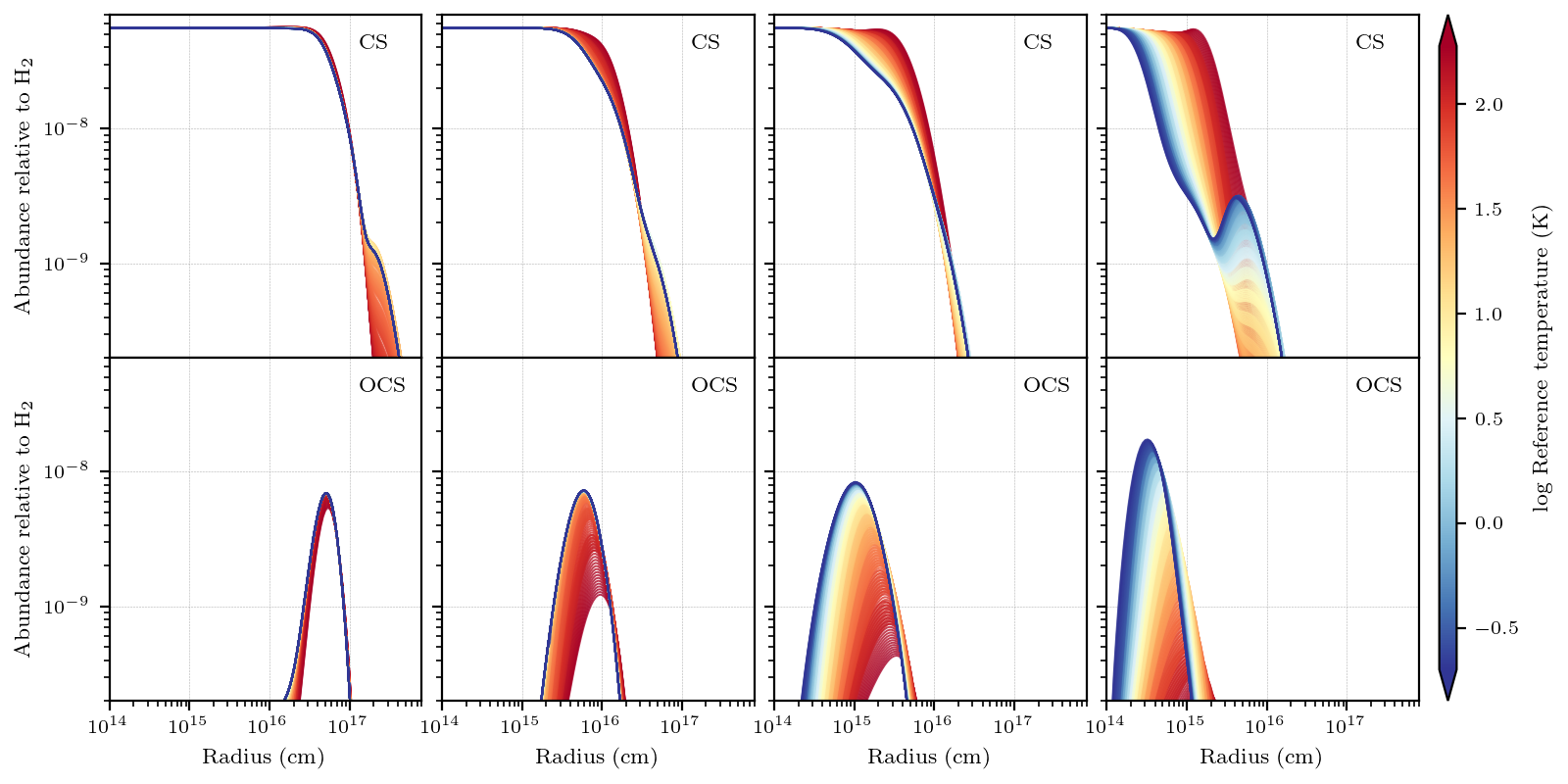}
		\caption{Fractional abundance profiles of CS and OCS in O-rich outflows, going from high to low outflow density. From left to right $(\vexp \ [{\rm \kms}],\Mdot \ [{\rm \Msolyr}])$: (25.0,\ $5\times10^{-5}$), (17.5,\ $2\times10^{-6}$), (10.0,\ $10^{-7}$), (2.5,\ $10^{-8}$).}
		\label{fig:O_CS_OCS}
	\end{figure*}
	\begin{figure*}
		\centering
		\includegraphics[width=0.92\textwidth]{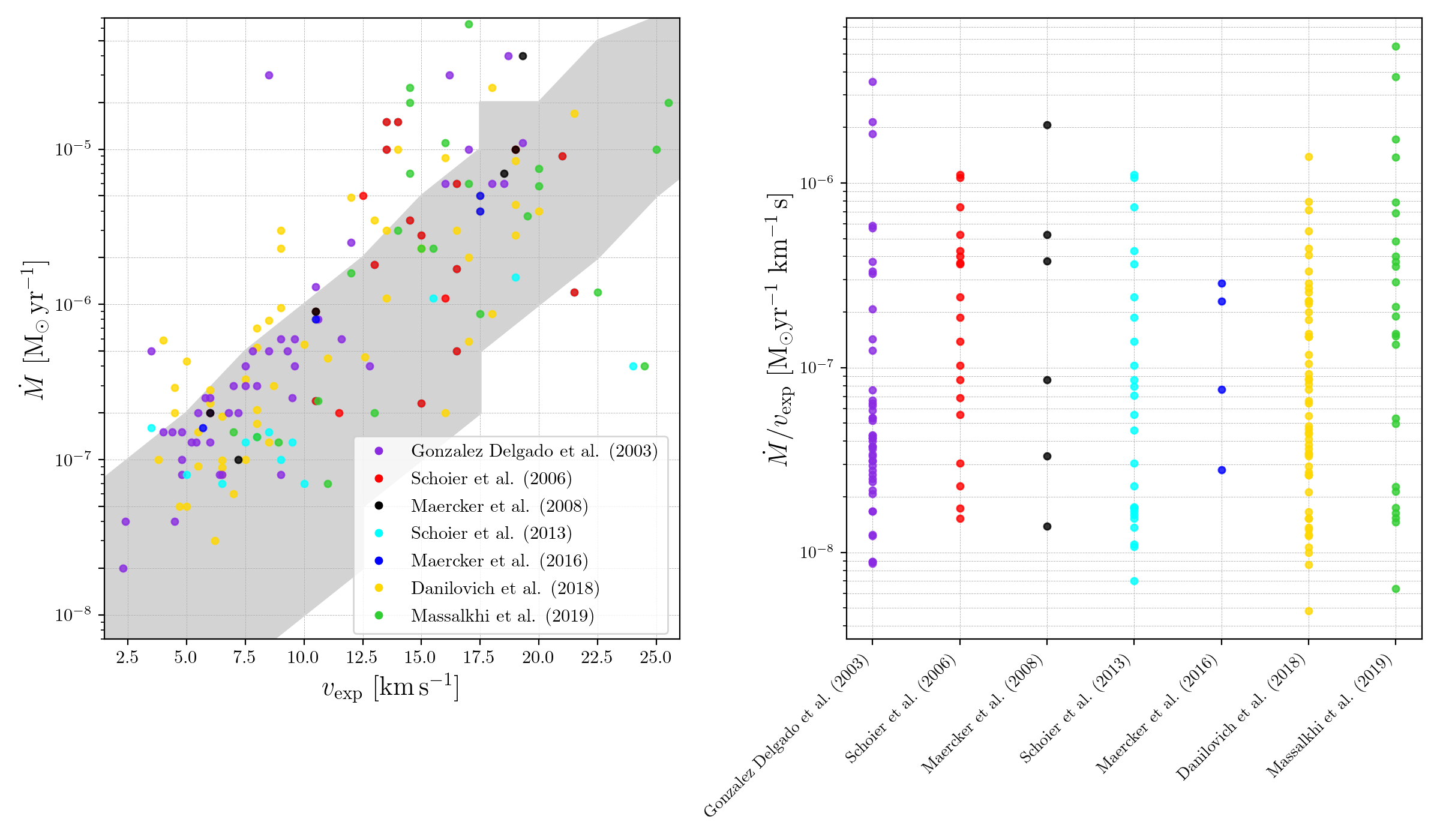}
		\caption{\textit{Left:} Visualisation of the ($\vexp,\Mdot$)-parameter space of our grid in shaded grey, see also Fig.\ \ref{fig:grid}. Literature sources indicated in circles: purple \protect\cite{Gonzalez2003}, red \protect\cite{Schoier2006}, black \protect\cite{Maercker2008}, cyan \protect\cite{Schoier2013}, blue \protect\cite{Maercker2016}, yellow \protect\cite{Danilovich2018}, and green \protect\cite{Massalkhi2019}. {\textit{Right}: Outflow density of the literature sources.}}
		\label{fig:literature_envs}
	\end{figure*}


	\bsp	
	\label{lastpage}
\end{document}